\documentclass[aps,pra,preprint,groupedaddress,showkeys,floatfix,showpacs]{revtex4}
\usepackage{bm}
\usepackage{setspace}
\usepackage{mathtools}
\usepackage{amsmath}

\begin{document}

\title{Dressing effects in laser-assisted ($e,2e$) process in fast electron-hydrogen 
atom collisions in an asymmetric coplanar scattering geometry }

\author{Gabriela Buic\u{a}}
\email{buica@spacescience.ro}
\affiliation{
Institute of Space Science, P.O. Box MG-36, Ro 77125,
Bucharest-M\u{a}gurele, Romania}

\begin{abstract}

We present the theoretical treatment of laser-assisted $(e, 2e)$ ionizing collisions 
in hydrogen for fast electrons, in the framework of the first-order Born approximation at 
moderate laser intensities and  photon energies beyond the soft-photon approximation.
The interaction of the laser field with the incident, scattered, and ejected electrons is 
treated nonperturbatively by using Gordon-Volkov wave functions,  while the atomic 
dressing is treated by using first-order perturbation theory.
Within this semi-perturbative formalism we obtain a new closed formula for the nonlinear 
triple differential cross section (TDCS), which is valid for linear as well circular 
polarizations. New analytical simple expressions of TDCS are derived in the weak field 
domain and low-photon energy limit.
It was found that for non-resonant $(e,2e)$ reactions the  analytical formulas
obtained for the atomic matrix element in the low-photon energy limit  give a good 
agreement, qualitative and quantitative,  with the numerical semi-perturbative model 
calculations.
We study the influence of the photon energy as well of the kinetic energy of the ejected 
electron on the TDCS, in the asymmetric coplanar geometry, and show that the dressing of 
the atomic target strongly influences the $(e, 2e)$ ionization process.
\end{abstract}
\date{\today}
\maketitle

\section{INTRODUCTION}
\label{I}

It is well known that the study of the atomic ionization process by collisions with 
electrons, the so-called $(e, 2e)$ reaction, reveals information about the 
electronic structure of the atomic target and residual ion \cite{b-j89},  and is of 
interest in collision theory or in other fields such as 
plasma physics or astrophysics, which need reliable scattering cross section data 
\cite{whelan93}.
Camilloni and coworkers \cite{camilloni72} were the first to use $(e,2e)$ reaction
as a tool for measuring the momentum distribution of the ejected electrons,
in a coplanar symmetric scattering geometry where the outgoing electrons have equal 
energies and polar angles, at high incident and outgoing electron energies.
Since then, an increasing number of $(e, 2e)$ experiments have been performed over the 
years for different target atoms and for various kinematical configurations, and the 
 electron momentum spectroscopy (EMS) has been developed to provide 
information on the electronic structure of atoms and molecules \cite{weigold99,smirnov99}.
The symmetric $(e,2e)$ reaction is the basis of EMS, also known as binary $(e,2e)$ 
spectroscopy, and is kinematically characterized by a large momentum transfer of the 
projectile electron and a small momentum  of the residual ion. 
Another useful scattering configuration is the coplanar asymmetric geometry with 
fast incident electrons (keV) and ejected electrons of low and moderate energies, where 
most of the $(e, 2e)$ reactions occur \cite{coplan94}.

In the past few decades the electron-impact ionization of an atom in the presence of a 
laser field has become increasingly interesting and it is often referred to as the 
\textit{laser-assisted $(e,2e)$ collision}  \cite{ehlotzky98}.
Recently,  H\"{\o}hr and coworkers \cite{hohr2007} performed the first 
kinematically-complete experiment for laser-assisted ionization in electron–helium 
collisions at high incident electron energy ($1$ keV) and showed significant differences 
of 
the triple differential cross section (TDCS) in comparison to the field-free 
cross-sections.
Very recently, Hiroi and coworkers \cite{hiroi2021} reported the observation the 
laser-assisted electron-impact ionization of Ar in an ultrashort intense laser field, and
showed that the signal intensity of the laser-assisted process for one-photon absorption 
obtained by integrating the signals over the detection angle ranges is about twice as 
large as that estimated by previous theoretical calculations in 
which the atomic dressing by the laser field is neglected \cite{cavaliere80-81}.

A large number of papers have been published so far and several theoretical 
approaches have been proposed, involving ejected electrons of low energies that are 
studied under the combined influence of the laser field and the Coulomb field of the 
residual ion.
The early theoretical works on laser-assisted $(e, 2e)$ scattering have neglected 
the dressing of the atomic target by the laser field or have used the closure 
approximation for laser-atom interaction.
First, Jain and Tzoar introduced the Coulomb-Volkov wave functions \cite{jain78}, which 
takes into account the influence of the Coulomb field of the nucleus on the final 
electron state.
Since then, the effect of the Coulomb interaction in the laser-assisted  $(e, 2e)$ 
collisions on hydrogen atom was studied in several papers by employing different types of 
final state wave functions like the Coulomb-Volkov or Coulomb corrected Gordon-Volkov 
wave functions. Banerji and Mittleman \cite{banerji81} calculated TDCS for ionization 
of hydrogen by electron impact, at low photon energies, in 
which the slow ejected electron was described by a modified Coulomb wave function,  and 
the laser-electrons interactions were included in the low-frequency approximation.
Cavaliere and coworkers  \cite{cavaliere80-81} studied laser-assisted $(e, 2e)$ 
collisions in hydrogen at low photon energies, high incident electron energies, and 
ejected electrons with moderate as well as small energies, in the first-order 
Born approximation, with the incident and scattered electrons  described by 
the Gordon-Volkov wave functions \cite{gordon,volkov}, while the ejected electron is 
represented by a modified Coulomb wave function.
Later on, the dressing of the atomic target by the laser field has been included in the
first-order time-dependent perturbation theory (TDPT), and therefore the influence of the 
laser parameters such as intensity, polarization, and photon energy has attracted a lot 
of interest from the theoretical point of view.
 Joachain and coworkers \cite{joachain88} extended the semi-perturbative theory 
of Byron and Joachain \cite{b-j}, and showed the strong influence of a laser field  on 
the dynamics of laser-assisted  $(e,2e) $ collisions in hydrogen, for fast incident and 
scattered electrons and slow ejected electrons, in the Ehrhardt
asymmetric coplanar geometry  \cite{ehrhardt}.
 For $(e,2e)$ collisions in hydrogen with slow ejected electrons, Martin and coworkers 
\cite{martin89} analyzed the influence of the laser parameters: photon energy, 
laser intensity, and polarization direction on the angular distribution of the  
ejected electrons.
The influence of laser polarization has also been discussed by Ta\"{\i}eb and coworkers
\cite{taieb1991}, who developed a dressed atomic  wave functions on a basis of Sturmian 
functions, which allowed to take into account accurately the contribution of the 
continuum spectrum to the dressing of the atomic states \cite{cionga93}. 
Very recently, Makhoute and coworkers \cite{ajana2019} presented their numerical results 
obtained for $(e, 2e)$ collision in atomic hydrogen in the symmetric and asymmetric 
coplanar scattering geometries, at large photon energies. For the direct scattering 
channel the calculation of the specific radial amplitudes was performed by expanding the 
atomic wave functions in a Sturmian basis, whereas the closure approximation was employed 
for the exchange channel.
\noindent
As mentioned before most of these previous theoretical works were focused on  scattering 
geometries involving slow ejected electrons, and only recently it was shown for ejected 
electrons of high energies that the laser field strongly modifies the $(e,2e)$ 
collisions.
New theoretical studies for laser-assisted EMS  at high impact energy and large 
momentum transfer were published and it was found that the atomic dressing, calculated 
in the closure- and low-frequency approximations, substantially influences the 
laser-assisted TDCSs at low  \cite{kouzakov2010,bulychev2012} and large photon energies 
\cite{khalil2017}.

The purpose of the present paper is to study the laser-assisted $(e, 2e)$ reactions in 
hydrogen, in which the target atom is ionized in collision with an electron beam 
in the presence of a laser field, for fast incident and outgoing electrons, in an 
asymmetric coplanar scattering geometry, beyond the soft-photon approximation.
We present a new method to derive the relevant atomic transition amplitude which takes 
into account the dressing of the target by the laser field.
The laser field alone cannot significantly ionize the hydrogen atom since the photon 
energy is considered below the ionization threshold and the laser intensity is not high 
enough to allow  ionization through a multiphoton process. 
We assume fast scattered electrons of sufficiently high velocity, such that we neglect 
their interaction with the Coulomb field of the remaining ion.
Similar to the approach used in the  Keldysh-Faisal-Reiss  approximation 
\cite{keldysh65,faisal73,reiss80}, the 
influence of the remaining ion on the final state of the fast ejected electron is 
neglected, since the residual Coulomb field is  weak compared to the laser field strength.
We follow the approach of Ref. \cite{joachain88} in which the semi-perturbative theory  
\cite{b-j} was generalized to laser-assisted fast $(e, 2e)$ collisions in atomic hydrogen.
In order to simplify the calculations we introduce several assumptions:
(a) It is reasonable to employ a first-order Born treatment of the projectile-atom 
interaction, since we consider fast nonrelativistic collisions such that the velocities 
of the projectile and outgoing electrons are much larger than the atomic unit
\cite{bransden,joa2012}.
(b) The non-relativistic Gordon-Volkov solutions are used for the incident and outgoing 
electrons to describe their interaction with the laser field.
(c) The laser field intensity is considered moderate, but much weaker than the atomic 
unit ($3.51 \times 10^{16} $ W/cm$^2$), in order to avoid direct one- and multiphoton 
ionization.  
In contrast to other theoretical works we take into account the atomic dressing effects 
in the first-order TDPT in the laser field, going beyond the soft-photon approximation.
The photon energy is considered below the ionization threshold of the hydrogen atom, 
and one-photon resonance transitions are allowed between the ground and excited states.
(d) Since the scattered and ejected electrons have high energies and of comparable order 
of magnitude, our semi-perturbative formalism takes into account the exchange effects 
in the first-order Born approximation.

\noindent
The manuscript is organized as follows.
In Sec. \ref{II} we present the theoretical method used in laser-assisted ionization of 
atomic hydrogen by electron impact, and derive  new analytical formulas for the 
ionization transition amplitudes and TDCSs by electron impact.
In the low-photon energy limit we  provide simple analytic formulas of TDCSs, in a closed 
form, for the laser-assisted $(e,2e)$ ionization process which include the atomic 
dressing effects.
Numerical results are presented in Sec. \ref{III}, where the TDCSs for laser-assisted 
electron impact ionization of hydrogen are analyzed as a function of the scattering angle 
of the ejected electron and as a function of the photon energy.
We have studied the modifications of the angular distributions of the ejected electrons 
due to the external laser field  at different ejected electron energies and photon 
energies.
Finally, summary and conclusions are given in  Sec. \ref{IV}.
Atomic units (a.u.) are employed throughout this manuscript, unless otherwise specified.

\section{Semi-perturbative theory}
\label{II}

The laser-assisted scattering of electrons by hydrogen atoms in  a laser field in which 
the atomic target is ionized, the so-called laser-assisted $(e, 2e)$ reaction, can be 
symbolically represented as:
\begin{eqnarray}
e^-(E_{i},\mathbf{k}_i) + {\rm H}(1s) +N_{i} \, \gamma (\omega,  \bm{\varepsilon})
\to  \nonumber \\
e^-(E_{f},\mathbf{k}_f) + e^-(E_{e},\mathbf{k}_e) +{\rm H}^+ + N_{f} \, \gamma (\omega,   
\bm{\varepsilon}),
\label{process}
\end{eqnarray}
\noindent
where $ E_{i} $ and $E_{f} $, and $\mathbf{k}_i \,(\theta_i,\varphi_i)$ and 
$\mathbf{k}_f \,(\theta_f,\varphi_f)$ represent the kinetic energy and the momentum 
vector of the incident and scattered projectile electrons, respectively, while
 $ E_{e} $ and $\mathbf{k}_e \,(\theta_e,\varphi_e)$ are the kinetic energy and  the 
momentum vector of the ejected electron, as plotted in Fig. \ref{fig1}.
Here $\gamma\,(\omega, \bm{\varepsilon})$ denotes a photon with the energy $\omega$ and 
the unit polarization vector $\bm{\varepsilon}$, and $N = N_{i}-N_{f}$   is the net 
number of exchanged photons between the projectile-atom scattering system and the laser 
field.
\noindent
The  laser field is treated classically, and within the dipole approximation is described 
as a monochromatic electric field,
\begin{equation}
\bm {{\cal E}} (t) =
(i/2){ \, \cal E}_{0} \, e^{-i  \omega  t} \, \bm{\varepsilon} + \rm{ c.c.},
\label{field}
\end{equation}
\noindent
where ${\cal E}_{0}$ represents the amplitude of the electric field.
The magnetic vector potential, $\bm{A}(t)$, is simply calculated from 
$ \bm {{\cal E}} (t) = -\partial_t \bm{A}(t) $, as
\begin{equation}
\bm{A}(t)  = ({ \cal E}_{0}/\omega)
\left[ \cos \omega t \cos (\xi/2) \,  \mathbf{e}_{j}
+    \sin \omega t \sin (\xi/2) \; \mathbf{e}_{l}
\right],
\label{vectpot}
\end{equation}
\noindent
where
$\bm{\varepsilon} =  \cos(\xi/2) \, \mathbf{e}_{j} +  i \sin(\xi/2) \, \mathbf{e}_{l}$
is the polarization vector of the laser beam, with $\mathbf{e}_{j}$ and $\mathbf{e}_{l}$ 
 two different unit vectors along different orthogonal directions.
  $\xi$ represents the degree of ellipticity of the laser field which varies in the 
range $  -\pi/2 \le \xi \le \pi/2$, and determines the ellipticity  of the field.
The value  $\xi=0$ corresponds to a linearly polarized (LP) laser field, while 
$\xi=\pi/2$ corresponds to a left-hand circularly polarized (CP) laser field.

\subsection{Laser-dressed electronic and atomic wave functions}

As mentioned before, we consider that the external laser field has a dominant influence
and neglect the Coulomb interaction between fast outgoing electrons and residual ion in 
the scattered and ejected electron wave functions \cite{keldysh65,faisal73,reiss80}.
At sufficiently high projectile kinetic energies, it is well known that the first-order 
Born approximation in the scattering potential can be used to describe the electron impact 
ionization process \cite{joachain88,bransden,joa2012}.
We assume fast incident and outgoing electrons with kinetic energies  much larger than the 
energy of a bound electron in the first Bohr orbit \cite{ehl1998}, 
since for the field-free $(e,2e)$ reaction in e-H collisions it is well known that the 
plane wave approximations agree well with experiment at kinetic energies above $200$ eV 
\cite{bransden,Lohmann81}.
Thus, in the non-relativistic regime, as long as both $E_f \gg 1 $ a.u. and $E_e \gg 
1 $ a.u., we describe the fast scattered and ejected electrons by Gordon-Volkov wave 
functions \cite{jain78,leone89,taj2004}.
We should mention that the use of a Coulomb-Volkov wave function provides a more accurate 
treatment at small impact kinetic energies, where the effect of the proton's potential on 
the incoming and outgoing electrons is important  \cite{whelan93,ehl1998,zhang2007}.
In order to avoid the direct one- and multiphoton ionization processes, we consider that 
the electric field amplitude is weak with respect to the atomic unit of electric 
field strengths, ${\cal E}_{0} \ll 5.1 \times 10^{9}$ V/cm, i.e. the strength of the 
laser field is much lower than the Coulomb field strength experienced by an electron in 
the first Bohr orbit.
\noindent
Therefore, we describe in a nonperturbative way  the initial and final states of the 
projectile electron, as well the final state of the ejected electron interacting with a 
laser field by  non-relativistic Gordon-Volkov wave functions \cite{gordon,volkov}, 
expressed in the velocity gauge as
\begin{equation}
 \chi_{ \mathbf{k}}^V  ({\mathbf{r}},t)=
    {(2\pi )^{-3/2}}  \exp \left[ { i{\mathbf{k}}\cdot {\mathbf{r}}
-i {\mathbf{k}} \cdot \bm{\alpha}(t)} -i E_k \, t - \frac{i}{2}  \int^t dt' \bm{A}^2(t') 
\right]
,
 \label{fe}
\end{equation}
where $\mathbf{r}$ and  $\mathbf{k}$ represent the position and momentum vectors, and 
$E_k  = k^2/2$ is the kinetic energy of the electron. 
$\bm{\alpha}(t) = \int^t dt' \bm{A}(t')$ describes the classical oscillation motion of 
a free electron in the electric field defined by Eq. (\ref{field}), and by using Eq. 
(\ref{vectpot}) we obtain
\begin{equation}
\bm{\alpha}(t)  = \alpha_{0}
\left[ \mathbf{e}_{j} \sin \omega t \cos (\xi/2) 
  + \, \mathbf{e}_{l} \cos \omega t \sin (\xi/2)
\right],
\label{quiver}
\end{equation}
where ${\alpha_{0}} = \sqrt{I}/ \omega^{2}$ is the amplitude of oscillation, and ${I}={ 
\cal E}_{0}^2$ denotes the laser intensity.
Obviously, as noticed from  Eq. (\ref{fe}), at moderate field strengths the largest 
effect of the laser field on the free-electron state is determined by a dimensionless 
parameter $ k \alpha_0 $, that depends on the electron and photon energies, and laser 
intensity. For example, a laser intensity of $1$ TW/cm$^2$, a photon energy  of 
$3.1$ eV, and an electron kinetic energy of $200$ eV result in a value of $k \alpha_0 
\simeq 1.58$, while the ponderomotive energy acquired by an electron in the electric 
field $U_{p}= {I}/ 4 \omega^{2}$ is about $0.015$ eV, and, therefore can be safely 
neglected compared to the photon and unbound electrons energies employed in the present 
paper.

The interaction of the hydrogen atom, initially in its ground state,  with a laser 
field at moderate field strengths is considered within the first-order TDPT.
An approximate solution for the wave function of an electron bound to a Coulomb potential 
in the presence of an electric field, also known as the dressed wave function, is 
written as
 \begin{equation}
\Psi_{1s}\left( \mathbf{r}_1, t\right)  = 
\left[
\psi_{1s}^{(0)} (\mathbf{r}_1,t)  + \psi_{1s}^{(1)}(\mathbf{r}_1,t)
\right]
\exp \left[-i { E}_{1}t -\frac{i}{2}  \int^t dt' \bm{A}^2(t')\right]
, \label{fat}
\end{equation}
where  $\mathbf{r}_1$ is the position vector of the bound electron, $ \psi_{1s}^{(0)} $ 
is the unperturbed wave function of the hydrogen atom ground state, and $\psi_{1s}^{(1)}$ 
represents the first-order perturbative correction to the atomic wave function due to the
external laser field.
We employ the following expression of the first-order correction in the velocity gauge, 
$\psi_{1s}^{(1)}$, as described by Florescu and Marian in Ref. \cite{vf1},
\begin{equation}
\psi_{1s}^{(1)}(\mathbf{r}_1,t) =-
\frac{ \alpha_0 \omega}{2} \left[
 \bm{\varepsilon} \cdot
 \mathbf{w}_{ 100}(E^{+}_1;\mathbf{r}_1 ) e^{-i\omega t} +
  \bm{\varepsilon}^* \cdot
 \mathbf{w}_{ 100}(E^{-}_1;\mathbf{r}_1 ) e^{i\omega t} \right]
,\label{co1m}
\end{equation}
with the  linear-response vector, ${\bf{w}}_{100}$, defined by
\begin{equation}
{\bf{w}}_{100}(E_1^{\pm} ;\mathbf{r}_1)
		 = -G_C(E_1^{\pm}) \, {\bf{P}} \, \psi_{1s}(\mathbf{r}_1),
\label{w}
\end{equation}
where $ \bf{P}$ denotes the momentum operator of the bound electron, and $G_C$ is the
Coulomb Green's function.
For the hydrogen atom in its ground state the linear-response vector was expressed in 
Ref. \cite{vf1} as
\begin{equation}
{\bf{w}}_{100}(E_1^{\pm} ;\mathbf{r}_1)
 = i (4 \pi)^{-1/2}{\cal B}_{101} (E_1^{\pm};{r}_1) \, \hat{\textbf{r}}_1
,
\label{wb}
\end{equation}
\noindent
where $\hat{\textbf{r}}_1=\textbf{r}_1/{r}_1 $, and the energies $E_1^{+}$ and 
$E_1^{-}$ take the following values 
\begin{equation}
E_1^{+} = {E}_1 + \omega +i 0,  \; \;  E_1^{-} = {E}_1 - \omega,
\label{Omega}
\end{equation}
 with $E_{1} =-13.6 $ eV representing the energy of the ground state.
The radial function ${\cal B}_{101}$ in  Eq. (\ref{wb})  was evaluated \cite{vf1} using 
the Schwinger's integral representation of the Coulomb Green's function in momentum space 
including both bound and continuum eigenstates, and can be expressed in terms of Humbert 
function, $\Phi_1$, as
\noindent
\begin{equation}
 {\cal B}_{101} (\tau;{r}_1)=\frac{2 \, \tau }{2-\tau}
 \left( \frac{2}{1+\tau}\right)^{2+\tau}  {r}_1 \, e^{-r_1/\tau}
 \Phi_1(2-\tau,-1-\tau,3-\tau, \xi_1, \eta_1),
\end{equation}
 where the parameter $\tau$ takes two values $\tau^{\pm}= 1/\sqrt{-2 E_1^{\pm}} $, and  
the variables of the Humbert function are $ \xi_1 = (1-\tau)/2$ and $ \eta_1 
=(1-\tau) {r}_1/\tau$.

\subsection{The nonlinear scattering matrix}
\label{scm}

We employ a semi-perturbative approach of the scattering process which is similar to that 
developed by Byron and Joachain \cite{b-j} for free-free transitions, in which the 
second-order Born correction is negligible compared to the laser-dressing effects.
The evaluation of the scattering amplitude is very challenging due to the 
complex three-body interaction: projectile electron, bound electron, and laser field.
However, since we assume that both scattered and ejected electrons have large kinetic 
energies the calculation simplifies, and we can derive a closed form expression for the 
TDCS.
Thus, the initial state of the scattering system is calculated as the product of the 
initial states of the fast incident electron and atomic target dressed by the laser field,
${\chi}_{ \mathbf{k}_i}^V( \mathbf{r}_0, t)$ and $ \Psi_{1s}(\mathbf{r}_1,t)$, while  
the final state  is calculated as the product of the final states of the fast scattered 
and ejected electrons, which are approximated as Gordon-Volkov wave functions.
Our treatment differs from that of  Ta\"{\i}eb and coworkers \cite{taieb1991} in the fact 
that we dress the fast projectile and ejected electrons to all order in the laser field 
and we dress the atomic target by using an atomic wave function corrected to the first 
order in the laser field \cite{vf1}.
As mentioned before,  we focus our study at moderate laser intensities ($I \le 
1$ TW/cm$^2$) and fast projectile electrons ($E_{i}$, $ E_{f}\ge 1$ keV) such that
the interaction between the projectile electron and hydrogen atom is well treated within 
the first-order Born approximation in the static scattering potential 
$V_{d}(r_0, r_1)=-1/r_0+ 1/{|\bf{r}_1-\bf{r}_0|}$ for the direct channel, and 
$V_{ex}(r_0, r_1)=-1/r_1+ 1/{|\bf{r}_1-\bf{r}_0|}$ for the exchange channel.

In order to describe the scattering process (\ref{process}) we employ the direct and 
exchange scattering matrix elements \cite{massey}, which are calculated at high 
kinetic energies of the projectile and ejected electrons as
\begin{eqnarray}
S_{fi,d}^{B1} & =& -i \int_{-\infty}^{+\infty} dt \,
\langle \chi_{\mathbf{k}_f}^V( \mathbf{r}_0,t)
\chi_{\mathbf{k}_e}^V(\mathbf{r}_1,t)  		 |V_d(r_0,{r}_1)|
{\chi}_{ \mathbf{k}_i}^V( \mathbf{r}_0, t) \Psi_{1s}(\mathbf{r}_1,t)
\rangle   \label{sm_d}, \\ 
S_{fi,ex}^{B1} & =& 
-i \int_{-\infty}^{+\infty} dt \,
\langle \chi_{\mathbf{k}_f}^V( \mathbf{r}_1,t)
\chi_{\mathbf{k}_e}^V(\mathbf{r}_0,t)  		 |V_{ex}(r_0,{r}_1)|
{\chi}_{ \mathbf{k}_i}^V( \mathbf{r}_0, t) \Psi_{1s}(\mathbf{r}_1,t)
\rangle 
,
\,
\label{sm_ex}
\end{eqnarray}
\noindent
 where $\chi_{\mathbf{k}_{i(f)}}^V$  and $\chi_{\mathbf{k}_e}^V$, given by Eq. (\ref{fe}), 
represent the Gordon-Volkov wave functions  of the projectile and emitted electrons 
embedded in the laser field,  whereas $\Psi_{1s} $, given by  Eq. (\ref{fat}),  
represents the wave function of the bound electron interacting with the laser field.
By using the Jacobi-Anger identity \cite{Watson},
$
 e^{-i x  \sin \omega t}  \equiv \sum_{N=-\infty}^{+\infty} J_N(x) e^{-i N \omega t},
$
we expand the oscillating part of the Gordon-Volkov wave functions occurring in the 
scattering  matrix  elements, Eqs. (\ref{sm_d}) and (\ref{sm_ex}), in terms of the 
ordinary Bessel functions of the first kind, $ J_N$, as
\noindent
\begin{equation}
\exp{[-i \,  \mathbf{q} \cdot \bm{\alpha}(t)]} =
\sum_{N=-\infty}^{+\infty} J_N({\cal R}_{q})  e^{-i N \omega t +i N \phi_{q}},
\label{bgv}
\end{equation}
\noindent
where the argument of the  Bessel function is defined by
$ {\cal R}_{q}= \alpha_{0}|\bm{\varepsilon} \cdot \mathbf{q} |$, 
and $\phi_{q}$ represents the dynamical phase which is calculated as
$ e^ {i \phi_{q}} = \bm{\varepsilon}\cdot \mathbf{q}/|\bm{\varepsilon}\cdot \mathbf{q}|$,
where  $ \mathbf{q} =  \mathbf{k_i} - \mathbf{k_f} - \mathbf{k_e} $ denotes the recoil 
momentum vector of the ionized target, $H^+$.
Clearly, for a CP laser field a change of helicity, i.e. $\bm{\varepsilon} \to 
\bm{\varepsilon}^*$, leads to a change of the sign of the dynamical phase, $\phi_{q} \to 
-\phi_{q}$ in the TDCS, while for a LP laser field $ e^ {i \phi_{q}} = \pm 1$, 
and $\phi_{q}=n \pi$ with $n$ an integer.

For the direct channel, by replacing Eqs. (\ref{fe}), (\ref{fat}), and (\ref{bgv}) into 
Eq. (\ref{sm_d}), we obtain the scattering matrix for electron-hydrogen collisions in a 
laser field, after performing the integration with respect to time,
\begin{equation}
S_{fi,d}^{B1} =- 2\pi i \sum_{N=N_{min}}^{+\infty}
\delta( E_{f} +E_{e}  -  E_{i} -   E_{1} - N \omega) \, T_{N,d}  \, ,
\label{smt}
\end{equation}
\noindent
where the  Dirac function, $\delta$, assures the energy conservation which implies that 
the kinetic energy of the scattered electron is determined by the relation
$  E_{f} = E_{i} +  E_{1} - E_{e} + N \omega $.
Here the kinetic energy of the residual ion, $E_{q} =q^2/2m_p$, has been 
neglected in comparison to any of the electrons kinetic energies, $E_j$, ($j=i, f$, and 
$e$), since the mass of the residual ion (proton) is much larger than the electron mass.
The energy spectrum of the scattered electron consists of an elastic line, $N=0$,
 and a number of sidebands corresponding to the positive and negative values  of $N$.
Obviously, for a given value of the ejected electron energy, $E_{e}$, the net number of 
exchanged photons is limited and cannot be smaller than a minimal value that is the 
integer of $N_{min} =( E_{e}  -E_{i} - E_{1}  )/\omega $.

The total nonlinear transition amplitude, $ T_{N,d} $, for the laser-assisted $ (e,2e)$ 
ionization  process in the direct channel can be split as a sum of two terms
\begin{equation}
T_{N,d} = T^{(0)}_{N,d} + T^{(1)}_{N,d} \, , 
\label{tgen}
\end{equation}
\noindent
where $T^{(0)}_{N}$ and $T^{(1)}_{N}$ represent the electronic and atomic transition 
amplitudes. The first term  on the right-hand side of the total transition amplitude  Eq. 
(\ref{tgen}), $T^{(0)}_{N}$, is the transition amplitude due to projectile 
electron contribution, in which the atomic dressing terms are neglected,
\begin{equation}
T^{(0)}_{N,d} = \frac{1}{2^{5/2}\pi^{7/2}}  \frac{ e^{ i N \phi_{q}}}{ \Delta^2}
 J_N({\cal R}_{q})  \int  d\mathbf{r}_1 \;
  e^{-i \mathbf{k}_e    \cdot \mathbf{r}_1 }
( e^{i  \mathbf{\Delta} \cdot \mathbf{r}_1 } - 	 1)
\psi_{1s} (\mathbf{r}_1)
,\label{tn0}
\end{equation}
\noindent
where the integration over the projectile coordinate, $\textbf{r}_0$, was performed using 
the Bethe integral, and $\mathbf{\Delta}= \mathbf{k_i} - \mathbf{k_f}$ is the vector of 
momentum transfer from the incident to the scattered electron.
After  performing the radial integration with respect to $\textbf{r}_1$ in Eq. 
(\ref{tn0}), the electronic transition amplitude can be simply expressed as
\begin{equation}
T^{(0)}_{N,d} =- \frac{1}{(2\pi)^2}
 J_N({\cal R}_{q}) \, f_{ion}^{B_1}(\Delta,q,k_e)  \, e^{ i N \phi_{q}},
\label{tne}
\end{equation}
 where
\begin{equation}
  f_{ion}^{B_1}(\Delta,q,k_e ) =   -\frac{2^{5/2} }{ \pi \Delta^2}
  \left[\frac{1}{(q^2+1)^2} - \frac{1}{(k_e^2+1)^2} \right],
  \label{tne-B}
\end{equation}
\noindent
is the direct scattering amplitude in the first-order plane-wave Born approximation for 
ionization of hydrogen atom by electron impact in the absence of the laser field 
\cite{b-j89,taj2004}.
In  the electronic transition amplitude, Eq. (\ref{tne}), 
the interaction between the laser field  and the projectile and ejected electrons is 
contained in the argument of the Bessel function 
$ {\cal R}_{q}=(\sqrt{I}/\omega^2)|\bm{\varepsilon} \cdot \mathbf{q} |$, 
and phase $ \phi_{q}$, being decoupled from the kinematic term.
This feature is a characteristic of employing Gordon-Volkov wave functions for fast 
electrons and moderate laser intensities \cite{ehl1998,acgabi2000,gabi2015-gabi2017}.
The field-free electronic scattering amplitude $f_{ion}^{B_1}$  
contains a factor, $-2/ \Delta^2$, which is related to the first-order Born amplitude 
corresponding to   scattering by the Coulomb potential ${-1}/{r_0}$, while the 
 two terms in the squared brackets of Eq. (\ref{tne-B}) are related to the 
momentum transfer to the residual ion, $q$, and the momentum of the ejected electron, 
$k_e$, respectively.
For the field-free $(e, 2e)$ collisions in the plane-wave Born approximation, the first 
term in the right-side-hand of  Eq. (\ref{tne-B}) gives rise to the so-called binary 
encounter peak \cite{weigold79}, which occurs at very low residual ion momentum $q \simeq 
0$.

The second term on the right-hand side of  Eq. (\ref{tgen}), $T^{(1)}_{N}$, represents 
the first-order atomic transition amplitude and corresponds to processes in which the 
hydrogen atom absorbs  or emits one photon and is subsequently ionized by the projectile 
electron impact.
$T^{(1)}_{N}$ occurs due to modification  of the atomic state by the laser field, the 
so-called \textit{atomic dressing},  which is described  by  the first-order radiative 
correction, $\psi^{(1)}_{1s}(\mathbf{r}_1,t)$, in Eq. (\ref{co1m}).
\noindent
After some straightforward algebra,  integrating over the projectile coordinate, 
$\textbf{r}_0$, the direct first-order atomic transition amplitude can be written as
\begin{equation}
T^{(1)}_{N,d} = - \frac{\alpha_{0} \omega}{2}
\left[ J_{N-1}({\cal R}_q)\;
{\cal M}_{at}^{(1)} ( \omega  ) e^{i (N-1)\phi_{q}}+
J_{N+1}({\cal R}_q)\;
{\cal M}_{at}^{(1)} ( -\omega ) e^{ i(N+1) \phi_{q}} 
\right]  ,
\label{tn1}
\end{equation}
\noindent
where  $ {\cal M}_{at}^{(1)}(\omega)$ denotes the specific first-order atomic transition 
matrix element related to one-photon absorption,
\begin{equation}
  {\cal M}_{at}^{(1)}(\omega) =  
  \frac{1}{2^{5/2}\pi^{7/2} \Delta^2}
 \int d \mathbf{r}_1 \;
  e^{-i \mathbf{k}_e \cdot \mathbf{r}_1 } \;(
  e^{ i \mathbf{\Delta} \cdot \mathbf{r}_1 } - 1)
	\bm{\varepsilon} \cdot  \mathbf{w}_{ 100}(E^{+}_1; \mathbf{r}_1),
 \label{mat}
\end{equation}
whereas the transition matrix element $ {\cal M}_{at}^{(1)} ( -\omega ) $ is related to 
one-photon emission
\begin{equation}
  {\cal M}_{at}^{(1)}(-\omega) =  
  \frac{1}{2^{5/2}\pi^{7/2} \Delta^2}
 \int d \mathbf{r}_1 \;
  e^{-i \mathbf{k}_e \cdot \mathbf{r}_1 } (
  e^{ i \mathbf{\Delta} \cdot \mathbf{r}_1 } - 1) \;
	\bm{\varepsilon}^* \cdot  \mathbf{w}_{ 100}(E^{-}_1; \mathbf{r}_1)
 ,\label{mat_em}
\end{equation}
where the energies $E_1^{\pm} $ are given in Eq. (\ref{Omega}).
Obviously, in Eq. (\ref{tn1}) only one photon is exchanged (emitted or absorbed)
between the laser field and the bound electron, while the remaining $N + 1$ or $N - 
1$  photons are exchanged between the laser field and the projectile electron.
By performing the radial integral over $\mathbf{r}_1 $ in Eq. (\ref{mat})  we derive the  
first-order atomic matrix element for one-photon absorption as,
\begin{equation}
{\cal M}_{at}^{(1)}(\omega) =
- \frac{1}{2^{3/2}\pi^{3} \Delta^2}
\left[
 (\bm{\varepsilon} \cdot \hat{\mathbf{q}})      \; {\cal J}_{101}(\omega, q ) -
 (\bm{\varepsilon} \cdot \hat{\mathbf{k}}_e )   \; {\cal J}_{101}(\omega, -k_e ) 
\right] ,
\label{mat+}
\end{equation}
while the following changes are made $ \omega \to -\omega$ and 
$\bm{\varepsilon} \to \bm{\varepsilon}^*$ in Eq. (\ref{mat+}) to obtain the first-order 
atomic transition matrix element, $ {\cal M}_{at}^{(1)} ( -\omega) $, for one-photon 
emission.
The expression of the  atomic radial integral ${\cal J}_{101}$, is given by
\begin{equation}
{\cal J}_{101}(\pm  \omega, p ) =  
{\int}_0^{\infty} dr_1 \; r_1^2 \; j_{1}(p r_1) \; {\cal B}_{101} ( E_1^{\pm}; r_1 ) 
, \label{jq}
\end{equation}
with ${\cal J}_{101}( \omega, -p ) = -{\cal J}_{101}( \omega, p )$, where $p=q$ or $k_e$.
After performing some algebra in Eq. (\ref{jq}), by using the expansion of the 
spherical Bessel function,  $j_{1}$, an analytical form of the radial integral is 
obtained in terms of two Appell's hypergeometric function, $F_1$, as
\begin{equation}
{\cal J}_{101}(\omega,p) =    \frac{2^{6} \tau} {p \, (2-\tau){(1+\tau)}^4} \;
Re\left[    a^3 F_1(b,1,3,b+1,x,y) - \frac{ia^2}{2p} F_1(b,2,2,b+1,x,y) \right] ,
\label{J101}
\end{equation}
 in which $ a={\tau}/{(1+i p \tau )}$, $b=2-\tau$, and the variables of the Appell's 
hypergeometric function are
\begin{equation}
x =\frac{\tau-1}{\tau+1},   \;\;
y =\frac{(1-\tau)(1-i p \tau)}{(1+\tau)(1+i p \tau)} ,
\end{equation}
where the parameter $\tau$ depends on the photon energy, $\omega$, and it takes two 
values, $\tau^{-} =1/ \sqrt{-2  E_1^{+ }}$ and $\tau^{+} =1/ \sqrt{-2  E_1^{- }}$, 
corresponding to the two energies $E_1^{+} $ and  $E_1^{-} $ defined in Eq. (\ref{Omega}).
\noindent
The  first-order atomic matrix element, Eq. (\ref{mat+}), has a structure that 
explicitly contains the scalar products $\bm{\varepsilon}\cdot  \hat {\mathbf{q}} $ and  
$\bm{\varepsilon} \cdot  \hat {\mathbf{k}}_e $, depends on the scattering geometry, 
being written in a closed form that allows us to analyze the dependence on the 
 laser field polarization.
The last term in the right-hand side of the electronic scattering amplitude and atomic 
matrix element, Eqs. (\ref{tne-B}) and (\ref{mat+}), occurs due to the non-orthogonality 
of the Gordon-Volkov wave function of the ejected electron and the initial ground-state 
wave function of the hydrogen atom. 
The structure of Eq. (\ref{mat+}) is also similar to other processes, with the vectors 
${\mathbf{q}}$ and $ \mathbf{k}_e$ replaced by  vectors which are specific to each 
particular process, such as elastic laser-assisted scattering of electrons by 
hydrogen atoms \cite{dubois86,acgabi2}, bremsstrahlung cross sections in the 
electron-hydrogen atom collisions \cite{dubois}, or laser-assisted electron-impact 
excitation of hydrogen atoms \cite{gabi2015-gabi2017}.

\subsection{The nonlinear scattering matrix for exchange scattering}
\label{scm_ex}

Our formalism does not neglect the exchange effects between the scattered and ejected 
electrons in both the  electronic and atomic terms, since fast incident and outgoing 
electrons are involved in the calculation, and, as in the EMS experiments their 
kinetic energies could have comparable orders of magnitude.
In the first-order Born approximation in the exchange potential, $V_{ex}$,  we obtain the 
exchange scattering matrix for the laser-assisted $(e,2e$) reactions, after performing 
the integration with respect to time in Eq. (\ref{sm_ex}),
\begin{equation}
S_{fi,ex}^{B1} =- 2\pi i \sum_{N=N_{min}}^{+\infty}
\delta( E_{f} +E_{e}  -  E_{i} -   E_{1} - N \omega) \, T_{N,ex} \, ,
\label{smt_ex}
\end{equation}
\noindent
where $ T_{N,ex} =T^{(0)}_{N,ex} +T^{(1)}_{N,ex} $.
The electronic transition amplitude for the exchange scattering, $T^{(0)}_{N,ex}$, in 
which the atomic dressing contribution is neglected, can be expressed as
\begin{equation}
T^{(0)}_{N,ex} =- \frac{1}{(2\pi)^2}
J_{N}({\cal R}_q) \, g_{ion,ex}^{B_1}(\Delta_e,q)\; e^{iN \phi_{q}}
, \label{tne_ex}
\end{equation}
where
\begin{equation}
g_{ion,ex}^{B_1}(\Delta_e,q) =  - \frac{2^{5/2}}{\pi \Delta_e^2(q^2+1)^2} ,
\label{tne-B_ex}
\end{equation}
denotes the electronic exchange amplitude in the absence of the laser field, that is in 
agreement to the Born-Ochkur approximation \cite{ochkur,massey}, and $\Delta_e$ 
represents the amplitude of the momentum transfer vector from the incident to the ejected 
electron, $\mathbf{\Delta}_e = \mathbf{k}_i- \mathbf{k}_e $.

\noindent
Similarly to the direct scattering, the first-order atomic transition amplitude for the 
exchange scattering can be expressed as
\begin{equation}
T^{(1)}_{N,ex} = - \frac{\alpha_{0} \omega}{2}
\left[ J_{N-1}({\cal R}_q)\;
{\cal M}_{at,ex}^{(1)} ( \omega  ) e^{i (N-1) \phi_{q}}+
J_{N+1}({\cal R}_q)\;
{\cal M}_{at,ex}^{(1)} ( -\omega ) e^{ i (N+1) \phi_{q}} 
\right] ,
\label{tn1_ex}
\end{equation}
\noindent
where
\begin{equation}
{\cal M}_{at,ex}^{(1)}(\pm \omega) =
- \frac{\bm{\varepsilon} \cdot \hat{\mathbf{q}}}{2^{3/2}\pi^{3} \Delta_e^2 }
 \;{\cal J}_{101}(\pm \omega, q ) 
 ,
\label{mat+_ex}
\end{equation}
 \noindent
 and $ {\cal J}_{101}(\pm\omega, q ) $ is calculated from Eq. (\ref{J101}).
 Obviously, the exchange effects for both electronic and atomic contributions to the 
transition amplitude vary like $\Delta_e^{-2}$, and cannot be neglected in comparison to 
the contribution of the direct scattering channel if ${k}_e $ and $  {k}_f $ are of 
comparable order of magnitude.
In contrast, for very fast incident and scattered electrons, with $k_{i}$  and  $k_{f} $ 
much larger than the atomic unit, and  slow ejected electrons,  $k_e \ll k_{f}$, the 
electronic and atomic exchange terms can be neglected compared to the corresponding direct 
terms.

\subsection{The low-photon energy approximation}
\label{scm_lim}

In the low-photon energy limit where the photon energy is  small compared to the 
ionization energy of the hydrogen atom (typically in the infrared region), it is worth 
 presenting some useful simple approximation formulas for the atomic transition amplitude.
In most theoretical works the analytical calculations cannot be done exactly and, as 
the photon energy remains small  is expected  that only few intermediate bound states to 
contribute to the atomic transition amplitude and thus to approximate the complicated 
analytical formulas.
This is the key of the closure approximation method \cite{b-j}, which consists in 
replacing the difference energy $E_n-E_1 $ by an average excitation energy 
$\bar\omega \simeq 
4/9$ a.u. for the hydrogen atom in approximating the sum over the intermediate states in 
the atomic transition amplitudes.
Here we present a different approach based on the low-frequency approximation (LFA), 
given by the lowest-order term of the expansion of the atomic matrix element ${\cal 
M}_{at}^{(1)} ( \omega  )$ in powers of the laser photon energy.
After some algebra we derive an approximate formula for the atomic radial integral, 
${\cal J}_{101}$, in the low-photon energy limit $\omega \ll |E_{1}|$ in  Eq. 
(\ref{J101}), in the first order in $\omega $,
\begin{equation}
{\cal J}_{101}(\omega,p) \simeq  - \frac{ 16 \, p}{(p^2 +1)^3}
 \left( 1 - \frac{\omega}{2} \, \frac{p^2 -9} {p^2 +1} \right),
 \label{J101_lim}
\end{equation}
where $p= q$ or $k_e$, and, therefore the atomic transition amplitude for the direct 
process, Eq. (\ref{tn1}), in the low-photon energy limit reads as
\begin{eqnarray}
T^{(1)}_{N,d} &\simeq &  {\alpha_{0} \omega}
\frac{2^{3/2}e^{i N\phi_{q}} }{\pi^{3} \Delta^2}
\left\{ J_{N-1}({\cal R}_q)  e^{-i \phi_{q}} 
\left[ \frac{  \bm{\varepsilon} \cdot\mathbf{q}}{(q^2 +1)^3} +
 \frac{  \bm{\varepsilon} \cdot\mathbf{k}_e}
 {(k_e^2 +1)^3} \right]\right.  \nonumber \\&&
 +\,   J_{N+1}({\cal R}_q) e^{i  \phi_{q}}
\left[ \frac{  \bm{\varepsilon}^* \cdot\mathbf{q}}{(q^2 +1)^3} +
 \frac{  \bm{\varepsilon}^* \cdot\mathbf{k}_e} {(k_e^2 +1)^3} \right] 
\nonumber  \\ &&
 + \,
\frac{\omega}{2} J_{N-1}({\cal R}_q)  e^{-i \phi_{q}}
\left[
 \frac{  \bm{\varepsilon} \cdot{ \mathbf{q}}\, (q^2-9)}{(q^2 +1)^4} +
  \frac{  \bm{\varepsilon} \cdot \mathbf{k}_e \,(k_e^2-9)}{(k_e^2 +1)^4} \right] 
\nonumber \\ &&  \left.
- \, 
\frac{\omega}{2} J_{N+1}({\cal R}_q)   e^{i \phi_{q}} 
\left[  \frac{ \bm{\varepsilon}^* \cdot{ \mathbf{q}} \,(q^2-9)}{(q^2 +1)^4} +
  \frac{ \bm{\varepsilon}^* \cdot\mathbf{k}_e \,( k_e^2-9)}{(k_e^2 +1)^4} \right] 
\right\}.
\label{tn1lim}
\end{eqnarray}

\noindent
For a LP laser field the following formula holds, 
$J_{N}({\cal R}_q) = J_{N}(\alpha_0 \, \bm{\varepsilon} \cdot{\mathbf{q}}) e^{-i 
N\phi_{q}} $, and  we obtain from Eqs. (\ref{tne}) and (\ref{tne-B}) the 
direct electronic transition amplitude, 
 \begin{equation}
T^{(0), LP}_{N,d} =
\frac{2^{1/2} }{\pi^{3} \Delta^2 }
J_N(\alpha_0 \, \bm{\varepsilon} \cdot{ \mathbf{q}})
  \left[\frac{1}{(q^2+1)^2} - \frac{1}{(k_e^2+1)^2} \right],
\label{tne_CA-LP}
\end{equation}
\noindent
and the direct atomic transition amplitude, Eq. (\ref{tn1lim}), simplifies to
\begin{eqnarray}
T^{(1), LP}_{N,d} &\simeq &  
\frac{2^{5/2} }{\pi^{3}}
\frac{  \omega}{\Delta^2}
\left\{ 
 N J_N(\alpha_0 \, \bm{\varepsilon} \cdot{ \mathbf{q}})
\left[ \frac{1}{(q^2 +1)^3} +
 \frac{\bm{\varepsilon} \cdot\mathbf{k}_e} {\bm{\varepsilon} \cdot \mathbf{q}}
  \frac{1}{(k_e^2 +1)^3} 
\right]  
 \right.  \nonumber \\&&
 +  \left. \, 
\frac{\alpha_{0}{\omega}}{2}
J_{N}^{'} (\alpha_0 \, \bm{\varepsilon} \cdot{ \mathbf{q}})\,  
\left[
 \frac{\bm{\varepsilon} \cdot { \mathbf{q}} \, (q^2-9)}{(q^2 +1)^4} +
  \frac{  \bm{\varepsilon} \cdot\mathbf{k}_e \,(k_e^2-9)}{(k_e^2 +1)^4} \right]
\right\},
\label{tn1lim-LP}
\end{eqnarray}
\noindent
where we have used the recurrence relation $ J_{N-1} (x) +J_{N+1} (x)=  J_{N}(x)(2N /x)$, 
and $ J_{N}^{'}$ is the first derivative of the Bessel function which satisfies the 
relation $ J_{N}^{'} (x) = [J_{N-1} (x) -J_{N+1} (x)]/2$, with $x= \alpha_0 ( 
\bm{\varepsilon} \cdot{ \mathbf{q}})$, \cite{Watson}. 
If we consider the lowest order in the photon energy $\omega$ in Eq. (\ref{tn1lim-LP})
\begin{equation}
T^{(1), LP}_{N,d} \simeq
\frac{2^{5/2} }{\pi^{3}}
\frac{ N\omega}{\Delta^2}
J_N(\alpha_0 \, \bm{\varepsilon} \cdot{ \mathbf{q}})
\left[ \frac{1}{(q^2 +1)^3} +
 \frac{\bm{\varepsilon} \cdot\mathbf{k}_e} {\bm{\varepsilon} \cdot \mathbf{q}}
 \frac{1 }{(k_e^2 +1)^3} \right] ,
\label{tn1lim-LP0}
\end{equation}
 we obtain the LFA formula for $N$-photon absorption atomic transition element  in the 
case of a LP field.
However, in the limit  $\omega \to 0$ for scattering parameters such that ${\cal 
R}_q \gg 1$, i.e there is a strong coupling between the projectile and ejected electrons 
and the laser field, the transition amplitudes derived  in the semi-perturbative approach 
do not diverge,  but approach the zero value, due to asymptotic behavior of the Bessel 
function of the first kind \cite{Watson} at large arguments,  
$J_N(x)\simeq \sqrt{2/\pi x} \cos (x-N \pi/2 -\pi/4)$, for $x \to \infty$.

Furthermore,  whenever the condition ${\cal R}_q \ll 1$  is satisfied, i.e. the 
perturbative regime of low laser intensities where $\alpha_0 \ll 1 $ a.u. and/or 
scattering kinematics with $|\bm{\varepsilon} \cdot{\mathbf{q}}|\ll 1 $ a.u., we can use 
the  approximate formula for the Bessel function at small arguments, 
\begin{equation}
J_N  ({\cal R}_q) \simeq 
\frac{1}{ N!} \left(\frac{{\cal R}_q}{2}\right)^N, 
 \quad \text{for  \;  $ N > 0  $},
\label{jnap}
\end{equation}
and $ J_{N}({\cal R}_q)= (-1)^{-N} J_{-N}({\cal R}_q) $, for $ N \le 0$  \cite{Watson}.
Thus, in the perturbative region with ${\cal R}_q \ll 1$, we  obtain from Eq. 
(\ref{tne_CA-LP}) a simple formula for the direct electronic transition amplitude for 
$N$-photon absorption ($N > 0$), as
\begin{equation}
T^{(0), LP}_{N,d} \simeq \alpha_{0}^{N}
\frac{2^{1/2} }{\pi^{3} \Delta^2 N !}
\left( \frac{\bm{\varepsilon} \cdot \mathbf{q} }{2}\right)^{N}
  \left[\frac{1}{(q^2+1)^2} - \frac{1}{(k_e^2+1)^2} \right],
\label{tne_CA}
 \end{equation}
 while for  the direct atomic transition amplitude for  $N$-photon absorption in 
the LFA we obtain from Eq. (\ref{tn1lim-LP0})
 \begin{equation}
 T^{(1), LP}_{N,d} \simeq  \alpha_{0}^{N}
\frac{2^{5/2} N \omega}{\pi^{3} \Delta^2 N ! }
\left( \frac{\bm{\varepsilon} \cdot \mathbf{q} }{2}\right)^{N}
\left[ \frac{1}{(q^2 +1)^3} +
 \frac{  \bm{\varepsilon} \cdot\mathbf{k}_e}{  \bm{\varepsilon} \cdot\mathbf{q}}
 \frac{1} {(k_e^2 +1)^3} \right].
 \label{tn1lim_CA}
\end{equation}
Expressions similar to Eqs. (\ref{tne_CA-LP})-(\ref{tn1lim-LP})  and 
Eqs. (\ref{tne_CA})-(\ref{tn1lim_CA}) can be easily derived for the exchange scattering 
channel.
As expected, the electronic and atomic transition amplitudes are important at scattering 
and ejected angles where the momenta $q$, $\Delta$, and $\Delta_e$  are small.
The ratio of the direct atomic and electronic transition amplitudes derived in the 
low-photon energy limit, Eqs. (\ref{tn1lim_CA}) and (\ref{tne_CA}), 
\begin{equation}
 \frac{T^{(1), LP}_{N,d}}{T^{(0), LP}_{N,d} } \simeq
 \frac{4N\omega}{q^2+1}
\left[1 + 
\frac{(q^2 +1)^3}{(k_e^2 +1)^3}   \frac{ \bm{\varepsilon} \cdot\mathbf{k}_e} { 
\bm{\varepsilon}\cdot\mathbf{q}} \right]
\left[ 1- \frac{(q^2+1)^2}{(k_e^2+1)^2} \right]^{-1},
\label{ratio_tne_CA}
\end{equation}
shows that, compared to the projectile electron contribution, the first-order atomic 
dressing effects for $\omega \ll 1 $ a.u. and laser parameters such that ${\cal R}_q \ll 
1$,  are increasing with the net number of exchanged photons, $N$, and photon energy, 
$\omega$, and are decreasing with the momenta of the ejected electron, $k_e$, and 
residual ion, $q$. 
Obviously, Eq. (\ref{ratio_tne_CA}) shows that differences occur in the TDCSs for 
absorption or emission of $N$ photons, which correspond to positive or negatives values 
of $N$, due to constructive or destructive interferences of the electronic and  atomic 
terms in TDCS.

In the case of one-photon absorption ($N=1$) in the perturbative regime with ${\cal R}_q 
\ll 1$ and low photon energies we can use the approximate formula for the Bessel 
function, 
Eq. (\ref{jnap}), and by keeping only the first order in laser field intensity, $I$, we 
obtain simple formulas for the direct electronic transition amplitude, Eq. (\ref{tne}),
\begin{equation}
T^{(0)}_{N= 1,d}\simeq 
\frac{   \sqrt{I} }{2^{1/2}\pi^3} 
\frac{ \bm{\varepsilon} \cdot \mathbf{q} }
{  \omega^2  \Delta^2}
  \left[\frac{1}{(q^2+1)^2} - \frac{1}{(k_e^2+1)^2} \right]
 , \label{t0lim}
\end{equation}
as well for the direct atomic transition amplitude derived  in the low-photon energy 
limit, Eq. (\ref{tn1lim}),
\begin{equation}
T^{(1)}_{N= 1,d} \simeq 
\frac{ 2^{3/2} }{\pi^{3}} 
\frac{\sqrt{I} }{\omega  \Delta^2}  
 \left[
 \frac{\bm{\varepsilon} \cdot \mathbf{q}}   {(q^2 +1)^3} 
 \left( 1  + \frac{\omega}{2} \;\frac{q^2-9}{q^2 +1}\right)
 +
 \frac{\bm{\varepsilon} \cdot \mathbf{k}_e} {(k_e^2 +1)^3} 
  \left( 1 +  \frac{\omega}{2} \;\frac{ k_e^2-9}{k_e^2 +1} \right)
 \right]
. \label{t1lim}
\end{equation}

\noindent
Moreover, if we keep the lowest order in the photon energy in Eq. (\ref{t1lim}) we 
obtain a quite simple formula for the direct atomic transition amplitude
\begin{equation}
T^{(1)}_{N= 1,d} \simeq 
\frac{ 2^{3/2} }{\pi^{3}} \frac{\sqrt{I} }{\omega  \Delta^2}  
 \left[
 \frac{\bm{\varepsilon} \cdot \mathbf{q}}   {(q^2 +1)^3}
 +
 \frac{\bm{\varepsilon} \cdot \mathbf{k}_e} {(k_e^2 +1)^3}
 \right], \label{t1lim1}
\end{equation}
in the LFA for one-photon absorption in the perturbative regime.

\noindent
Similarly, for the  exchange scattering we derive simple approximate formulas for the 
electronic and atomic transition amplitudes at ${\cal R}_q \ll 1$, Eqs. (\ref{tne_ex}) 
and (\ref{tn1_ex}) in the low-photon energy limit, as
\begin{equation}
T^{(0)}_{N= 1,ex}\simeq 
\frac{ 1 }{2^{1/2}\pi^3} 
\frac{  \sqrt{I} } {  \omega^2  \Delta_e^2} 
\frac{\bm{\varepsilon} \cdot \mathbf{q}}{(q^2+1)^2} 
 , \label{t0lim_ex}
\end{equation}
\begin{equation}
T^{(1)}_{N= 1,ex} \simeq 
\frac{ 2^{3/2}   }{\pi^{3}} 
\frac{\sqrt{I}}{\omega  \Delta_e^2} 
 \frac{\bm{\varepsilon} \cdot \mathbf{q}}   {(q^2 +1)^3} 
 \left( 1  + \frac{\omega}{2} \;\frac{q^2-9}{q^2 +1}\right).
 \label{t1lim_ex}
\end{equation}
The infrared divergence in the limit  $\omega \to 0$ is evident in all  the above
electronic and atomic transition amplitude expressions derived at ${\cal R}_q \ll 1$.
Thus, in the perturbative regime and low-photon energy approximation the electronic 
transition amplitude varies like $\omega^{-2}$,  while the  atomic transition amplitude 
varies  like $\omega^{-1}$, which is reminiscent of the infrared divergence of quantum 
electrodynamics  \cite{Jauch} and Low theorem \cite{Low} in  the  limit $\omega \to 0$.
Clearly, these simple analytical formulas we have derived for one-photon absorption, as 
well  for nonlinear atomic transition amplitudes might provide more physical insight into 
the laser-assisted $(e,2e)$ reactions.

\subsection{The triple differential cross section}
\label{tdcs_ch}

It is well known that the TDCS can provide useful information about collision dynamics 
in the electron-impact ionization process \cite{ehlotzky98}. For laser-assisted $(e,2e)$ 
collisions accompanied by the transfer of $N$ photons, we calculate the nonlinear TDCS in 
the first-order Born approximation in the scattering potential, for unpolarized incident 
projectile and hydrogen beams, and without distinguishing between the final spin states of 
the electrons,
\begin{equation}
\frac{d^3{\sigma}_{N}^{B1}}{ d\Omega_{f} \, d\Omega_{e} \, d E_{f}} =
 {(2\pi)}^4  \frac{k_f k_e}{k_i} 
 \left( \frac 1 4 {\left| T_{N,d} + T_{N,ex} \right|}^2 +  \frac 3 4 
{\left| T_{N,d} - T_{N,ex}\right|}^2 \right) 
 , \label{tdcs}
\end{equation}
\noindent
averaged over the initial spin states and summed over the final spin states.
The projectile  electrons are scattered  into the solid angle $\Omega_{f}$ and 
$\Omega_{f} +d\Omega_{f}$  with the kinetic energy between $E_f$ and $E_f+dE_f $, and the 
ejected electrons are emitted within the solid angle $\Omega_{e}$ and $\Omega_{e} 
+d\Omega_{e}$.
The TDCS is a function of the electrons momentum vectors ${\mathbf k}_i, {\mathbf k}_f$, 
and  ${\mathbf k}_e$, and depends on the laser parameters: intensity $I$, photon energy 
$\omega$, and polarization $\bm{\varepsilon}$. 
The dominant contribution to TDCS is due to collisions involving small momentum transfers 
$\Delta$ and $\Delta_e$, small momentum of the residual ion $q$, or near resonance 
photon energies.
The  TDCS for the laser-assisted $(e,2e)$ process is given by
\begin{equation}
\frac{d^3{\sigma}^{B1}}{ d\Omega_{f} \,  d\Omega_{e} \,  d E_{f}} =
\sum_{N=N_{min}}^{+\infty}
  \frac{d^3{\sigma}_{N}^{B1}}{ d\Omega_{f} \, d\Omega_{e} \, d E_{f}} \,.
\label{tdcs-N}
\end{equation}
By integrating TDCS over the direction of the scattered electrons, $\Omega_{f} $, we 
obtain the double differential cross section  of the ejected electrons, while by 
integrating TDCS over the direction of the ejected electrons, $\Omega_{e} $, we  derive 
the double differential cross section of the scattered electrons. Finally, the total 
ionization cross section is deduced by integrating over the angles and energies of 
the scattered and ejected electrons.

By neglecting the atomic dressing in Eq. (\ref{tdcs}), namely $T_{N,d}^{(1)} \simeq 0$ 
and  $T_{N,ex}^{(1)} \simeq 0$, at small momentum of the residual ion, $q 
\ll k_e$, we obtain a simple formula for  the laser-assisted TDCS,
 \begin{equation}
\frac{d^3{\sigma}_{N}^{B1}}{ d\Omega_{f} \,  d\Omega_{e} \,  d E_{f}} \simeq
\frac{k_f k_e}{k_i} 
  \, |J_{N}({\cal R}_q)|^2
   \frac{4}{ \Delta^4 } 
 \left( 1  - \frac{\Delta^2}{\Delta_e^2}  +  \frac{\Delta^4}{\Delta_e^4}
 \right)  |\psi_{1s}^{(0)}(q)|^2  
 , \label{tdcs-mott}
\end{equation}
\noindent
that ``decouples'' into a product of three factors: (i) the squared Bessel function which 
includes the laser-projectile and ejected electrons interaction, (ii)
the electron-electron  collision factor  in the first-order Born approximation 
$$ f_{ee}^{B1} =  \frac{1}{4\pi^4 \Delta^4} \left( 1  - \frac{\Delta^2}{\Delta_e^2} +  
\frac{\Delta^4}{\Delta_e^4}  \right), $$
 that is the absolute square of the half-off-shell Coulomb-matrix element summed and 
averaged over final and initial spin states for fast projectile and outgoing electrons 
\cite{takahashi2006},
and (iii) $|\psi_{1s}^{(0)}(q)|^2 ={8 }\pi^{-2} (q^2+1)^{-4}$ that represents the squared 
momentum-space wave function for the ground state of atomic hydrogen \cite{b-j89}.
Equation (\ref{tdcs-mott}) is in agreement to the TDCS derived for EMS  by Kouzakov and 
coworkers, namely Eq. (26) in Ref. \cite{kouzakov2010}.
The half-off-shell Mott scattering TDCS, for fast projectile and outgoing electrons that 
includes the exchange terms, \cite{massey,weigold99}, is simply calculated as 
$(2\pi)^4f_{ee}^{B1}$,
 \begin{equation}
\left( \frac{d{\sigma}}{ d\Omega_{e} }\right)_{ee} =   \frac{4}{ \Delta^4 } 
 \left( 1  - \frac{\Delta^2}{\Delta_e^2} +  \frac{\Delta^4}{\Delta_e^4} 
 \right)   
 . \label{tdcs-mott-ee}
\end{equation}
\noindent
If we take into account the atomic dressing in Eq. (\ref{tdcs}) in the low-photon energy 
limit $\omega \ll |E_1|$, and consider the lowest order in the photon energy in Eq. 
(\ref{tn1lim}), at small momentum of the residual ion, $q \ll k_e$, we obtain
 \begin{equation}
\frac{d^3{\sigma}_{N}^{B1}}{ d\Omega_{f}  \, d\Omega_{e}  \, d E_{f}} \simeq
\frac{k_f k_e}{k_i} 
  \, |J_{N}({\cal R}_q)|^2
  \left( \frac{d{\sigma}}{ d\Omega_{e} }\right)_{ee}
   \left(1+    \frac{4 N \omega }{q^2+1}\right) ^2 |\psi_{1s}^{(0)}(q)|^2
 , \label{tdcs-mott-1}
\end{equation}
that is in agreement to the laser-assisted TDCS derived in the low-photon energy  
approximation for EMS by Bulychev and coworkers, namely Eqs. (9)-(11) in Ref. 
\cite{bulychev2012}.
In contrast to Eq. (\ref{tdcs-mott}) in which the atomic dressing effects are neglected, 
the TDCS Eq. (\ref{tdcs-mott-1}) does not obey the well-known Kroll-Watson sum rule 
\cite{k-w}.
Obviously, the TDCS in the laser-assisted $(e,2e)$ collisions provides valuable 
information about the collision dynamics \cite{weigold99}, electronic structure of the 
target, and can be used to derive the momentum density distribution of the target 
electron, which was first demonstrated for hydrogen and helium atoms 
\cite{Lohmann81,b-j89}.

\section{NUMERICAL EXAMPLES AND DISCUSSION}
\label{III}

\begin{figure}
\centering
\includegraphics[width=2.20in,angle=0]{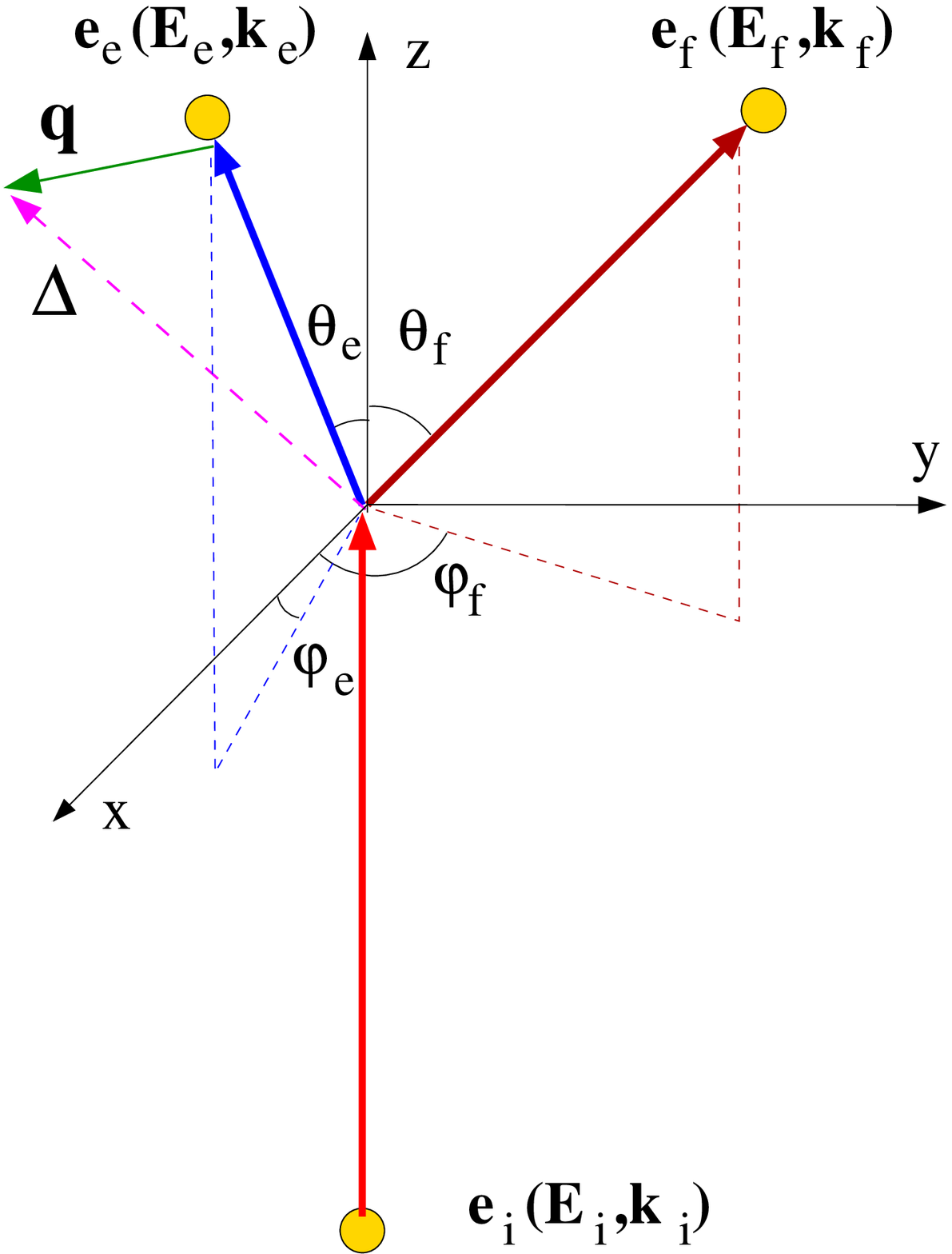}
\caption{(Color online) 
 Schematic representation of the scattering geometry for the $(e,2e)$ reaction. 
 The incident electron has energy $E_i$ and momentum $\mathbf{k_i}$, while  the 
scattered electron has energy $E_f$ and momentum $ \mathbf{k_f}$, and is detected at a 
fixed scattering angle $ \theta_f$.
The ejected electron has energy $E_e$ and momentum $ \mathbf{k_e}$, and its detection 
angle $ \theta_e$ is varied.
$\mathbf{\Delta}= \mathbf{k_i}-\mathbf{k_f}$ denotes the momentum transfer vector from 
the incident to the scattered electron, and $\mathbf{q}= \mathbf{\Delta}-\mathbf{k_e}$ is 
ion recoil momentum vector.
For a coplanar scattering geometry lying in the $(x, z)$-plane the azimuthal angles  
$\varphi_f $ and $ \varphi_e$ are equal to $0^{\circ} $ or $180^{\circ} $,  with the 
incident electron propagating in the $z$-axis direction, $ \varphi_i = 0^{\circ}$ and $ 
\theta_i = 0^{\circ}$.
}
\label{fig1}
\end{figure}

In this section we present our numerical results for the laser-assisted electron-impact 
ionizing collisions in hydrogen, described by Eq. (\ref{process}), for 
fast incident and outgoing electrons, and we apply the semi-perturbative formulas derived 
in Sec. \ref{II} to calculate the nonlinear TDCSs  in the presence of a LP 
laser field. 
Obviously, due to the complicated analytical form of the laser-dressed atomic wave 
function, the total scattering amplitude has to be numerically evaluated.
It is worth pointing out that the electronic and atomic transition amplitudes, Eqs.  
(\ref{tne}), (\ref{tn1}), (\ref{tne_ex}), and (\ref{tn1_ex}), as well their 
approximations derived in Subsec. \ref{scm_lim} are applicable for arbitrary scattering 
configurations and laser field polarizations.
\noindent
We study the laser-assisted $(e,2e)$ process in the coplanar geometry depicted in Fig. 
\ref{fig1}, in which the momenta of the electrons, $\mathbf{k}_i, \mathbf{k}_f$, and 
$\mathbf{k}_e$, lie in the same plane where the two outgoing electrons are detected in 
coincidence at the scattering angles $\theta_f$  and  $\theta_e$, with equal 
corresponding azimuthal angles $\varphi_f=\varphi_e=\varphi_i$.
The momentum vector of the incident electron, $\mathbf{k}_i$, is taken parallel to the 
$z$ axis, with $\theta_i=0^{\circ}$ and $\varphi_i=0^{\circ}$, and the scattering angle 
$\theta_f$  of the scattered electron is fixed, while the angle $\theta_e$ of the ejected 
electron is varied.
The asymmetric scattering geometry is considered in which $\theta_f \neq\theta_e$ and 
${k}_f \neq {k}_e$. At this point it is useful to recall the differences between 
the symmetric and asymmetric scattering geometries, namely the symmetric geometry is 
defined by the requirement that the scattering angles and energies of the scattered and 
ejected electrons are equal.
In a kinematically complete experiment by measuring the momentum vectors of both ejected 
electron and ionized target, $\mathbf{k}_e$ and $ \mathbf{q}$, we can deduce the 
momentum of the scattered electron, $\mathbf{k}_f = \mathbf{k}_i - \mathbf{k}_e - 
\mathbf{q}$, as well as the momentum transfer of the scattered electron, $\mathbf \Delta = 
\mathbf{k}_i- \mathbf{k}_f $,  occurring during the collision \cite{hohr2007}.
\noindent
Thus, from the energy conservation law, the final momentum of the projectile is given by
${k_f}  = {( k_i^{2} - k_e^{2}+2E_1  + 2N \omega) }^{1/2}$, while the momentum transfer of 
the projectile is simple calculated as  
$\Delta =  ( k_i^{2} +  k_f^{2} -2  k_i k_f \cos \theta_f )^{1/2}$.
The Cartesian components of  the momentum transfer vector, $ \mathbf{\Delta} $, are given 
by $(-k_f \sin \theta_f, 0, k_i -k_f \cos \theta_f )$ and the amplitude   $\Delta $ 
varies in the range $ |k_i -k_f| \leq \Delta \leq   k_i +k_f$, for forward 
$\theta_f=0^{\circ}$ and backward $\theta_f=180^{\circ}$ scattering, respectively. 
Similarly, the amplitude of the momentum transfer vector $ \mathbf{\Delta}_e $ is 
calculated as $\Delta_e =  ( k_i^{2} +  k_e^{2} -2  k_i k_e \cos \theta_e )^{1/2}$.
The amplitude of the recoil momentum vector of the residual ion, ${q}$, is given by
\begin{equation}
 q =  {[  \Delta^{2}  +  k_e^{2} -2 k_i k_e \cos \theta_e  + 2 k_f k_e \cos (\theta_f - 
\theta_e ) ]}^{1/2}.
\label{q}
\end{equation}

\noindent
The argument ${\cal R}_q$ of the  Bessel functions is calculated as
\begin{equation}
{\cal R}_q^2=   {\cal R}_{i}^{2} +  {\cal R}_{f}^{2} +  {\cal R}_{e}^{2}
- 2 {\cal R}_{i} {\cal R}_{f} \cos  (\phi_i- \phi_f)  
- 2 {\cal R}_{i} {\cal R}_{e}  \cos (\phi_i- \phi_e) 
+ 2 {\cal R}_{f} {\cal R}_{e} \cos (\phi_f - \phi_e ) ,
\nonumber
\end{equation}
where  $ {\cal R}_{s}= \alpha_{0}|\bm{\varepsilon} \cdot \mathbf{k}_s |$ and 
 $ e^{i \phi_{s}} = \bm{\varepsilon}\cdot \mathbf{k}_s / |\bm{\varepsilon}\cdot 
\mathbf{k}_s |$,  with $s=i,f$, and $e$.
For a LP laser field the dependence of $ {\cal R}_{s}$ on the laser polarization is given 
by
$ {\cal R}_{s}= \alpha_{0} |\mathbf{e}_{j}\cdot \mathbf{k}_s |$ and $\phi_{s}= n\pi$, 
while for a CP field with the  polarization unit vector
$\bm{\varepsilon}= ( \mathbf{e}_{j} +  i \mathbf{e}_{l} ) /\sqrt{2}$,
we obtain
$ {\cal R}_{s}= (\alpha_{0}/\sqrt{2})\sqrt{(\mathbf{e}_j\cdot \mathbf{k}_s )^2
              +(\mathbf{e}_l\cdot \mathbf{k}_s )^2}$ and
$\phi_{s}=\arctan{(\mathbf{e}_l\cdot \mathbf{k}_s )/(\mathbf{e}_j\cdot \mathbf{k}_s )} 
+n\pi$,  where $n$ is an integer.
In our numerical calculation we consider that the laser field is linearly polarized in 
the same direction along the momentum vector of the incident electron, 
$\bm{\varepsilon} || \mathbf{k}_i$.
Specifically, for a LP laser field and a coplanar scattering geometry with $\phi_i = 
\phi_f =\phi_e =0^{\circ}$  the argument of the  Bessel function simplifies to
$ {\cal R}_q=  \alpha_0 ( k_i - k_f \cos \theta_f -k_e \cos \theta_e )^{1/2}$.

To start with, we have checked that the numerical results of TDCSs for the ($e, 2e$) 
scattering of fast electrons by hydrogen atoms in their ground state are in agreement with 
earlier numerical data published in the literature.
A very good agreement is obtained  with the numerical results of TDCS for one- and 
two-photon exchange presented in Fig. 1 of  Ref. \cite{kouzakov2010} and Figs. 1 and 2 of 
Ref. \cite{bulychev2012}, under the kinematical conditions of EMS (small momentum of the 
residual ion $q$ and large momentum transfers $\Delta$ and $\Delta_e$), for incident  
electrons of kinetic energy $E_i=2013.6 $ eV, in a noncoplanar symmetric scattering 
geometry, and a LP laser of intensity $4 \times 10^{12}$ W/cm$^2$, calculated in the 
low-frequency approximation  at $ \omega =1.17$ eV.
At an incident electron kinetic energy $E_i=500 $ eV, in a coplanar symmetric 
geometry,  $\bm{\varepsilon} \| \bm \Delta$,  and a LP laser of intensities $I=1.3 \times 
10^7$ W/cm$^2$, $10^2 \times I, 10^4 \times I$, and $ 10^6 \times I$, the  behavior of the 
TDCS calculated from Eq. (\ref{tdcs}) is in fair agreement, up to a scaling factor, to the 
first-order Born calculation of TDCS for the ionization of hydrogen shown in Figs. 2 and 3 
 of Ref. \cite{khalil2017}.
Since  the atomic wave function was calculated  within  the closure 
approximation \cite{khalil2017}, our numerical results disagree at larger photon energies 
$\omega>3$ eV where the atomic dressing effect is more important, and cannot be accurately 
described by this approximation.
At the resonance photon energy of $ 10.2 $ eV, laser intensity of $1.3 \times 10^7 
$ W/cm$^2$, and polarization $\bm{\varepsilon}||\bm k_i$, in the Ehrhardt asymmetric 
coplanar geometry, with the incident and ejected electrons kinetic energies $E_i=250$ eV 
and $E_e=5$ eV, and scattering angle $ \theta_f= 3^{\circ}$, the TDCS given by Eq. 
(\ref{tdcs}) is in an satisfactory agreement with the first-Born TDCS for the ionization 
of  hydrogen plotted in Fig. 3(a) of Ref. \cite{taieb1991} where the atomic wave function 
is calculated using a Coulomb-Sturmian basis. Obviously, despite the low value of the 
ejected electron kinetic energy, the agreement is due to the fact that for one-photon 
resonance the TDCS is dominated by the atomic contribution due to $1s-2p$ excitation.

Now, we return our discussion to the scattering geometry depicted in Fig. \ref{fig1} 
where the laser polarization, $\bm{\varepsilon}$,  is parallel to the incident 
electron momentum direction, $ \mathbf{k}_i$, and the outgoing electrons move 
asymmetrically with respect to the direction of the incident electron, with different 
scattering and ejected angles, and different kinetic energies.
We have chosen high kinetic energies  of the projectile and ejected electrons (compared 
to the atomic scale), moderate laser intensities below $1$ TW/cm$^2$ which correspond to  
electric field strengths lower than $2.7 \times 10^7$ V/cm, and have considered photon 
energies below the ionization threshold of the  hydrogen atom.
Specifically, a laser intensity of $1 $ TW/cm$^2$ and a photon energy of $1.55$ eV 
(Ti:sapphire laser) result in a quiver motion amplitude $\alpha_{0} \simeq 1.64$ a.u. and 
an argument of the ordinary Bessel function  
${\cal R}_q \simeq  1.64|\bm{\varepsilon}\cdot {\mathbf{q}}|$,  
while for a larger photon energy of $3.1$ eV (Ti:sapphire second harmonic) the 
corresponding amplitude $\alpha_{0}$ and the argument ${\cal R}_q $ are about $4$ times 
smaller.
The numerical results obtained for TDSCs in the first-order Born approximation in the 
scattering potential,  Eq. (\ref{tdcs}), are compared with those obtained by considering 
the atomic contribution in the LFA,  Eq. (\ref{tn1lim-LP0}),  and those obtained  by 
neglecting the dressing of the target by setting $T_{N,d}^{(1)} \simeq 0$ and 
$T_{N,ex}^{(1)} \simeq 0$ in Eq. (\ref{tdcs}).

\begin{figure}
\centering
\includegraphics[width=3.5in,angle=0]{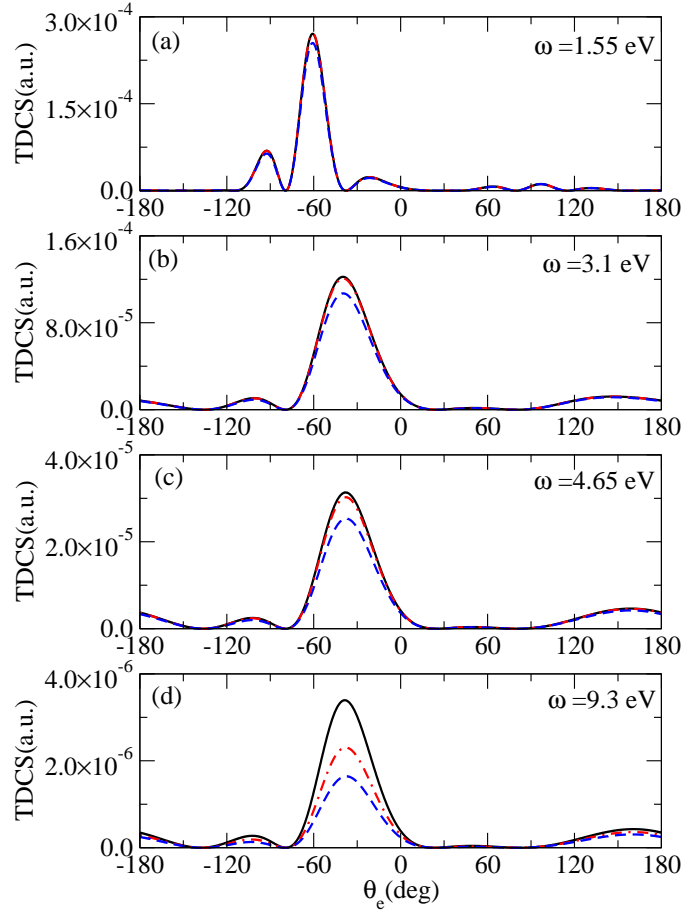}
\caption{(Color online)
The TDCSs, in atomic units, for the ionization of hydrogen from the ground state by 
electron impact in the presence of a LP laser field with $\bm{\varepsilon} \| \bm 
k_i$, as a function of the ejected electron angle $\theta_e$, for one-photon absorption.
The kinetic energies of the incident and ejected electron are $E_i = 2$ keV and $E_e=200$ 
eV, and the scattering angle is $\theta_f = 5^{\circ}$.
The laser intensity is $I=1$ TW/cm$^2$, while the photon energy is $\omega=1.55 $ eV in 
 Fig. \ref{fig2}(a), $ 3.1$ eV in Fig. \ref{fig2}(b), $4.65 $ eV in Fig. \ref{fig2}(c),  
and $9.3$ eV in Fig. \ref{fig2}(d). 
The solid lines correspond to the laser-assisted TDCSs calculated from Eq. (\ref{tdcs}), 
the dashed lines correspond to TDCSs in which the atomic dressing is neglected, and 
dot-dashed lines correspond to TDCSs in which the atomic dressing is considered in the 
LFA.
}
\label{fig2}
\end{figure}

\begin{figure}
\centering
\includegraphics[width=3.5in,angle=0]{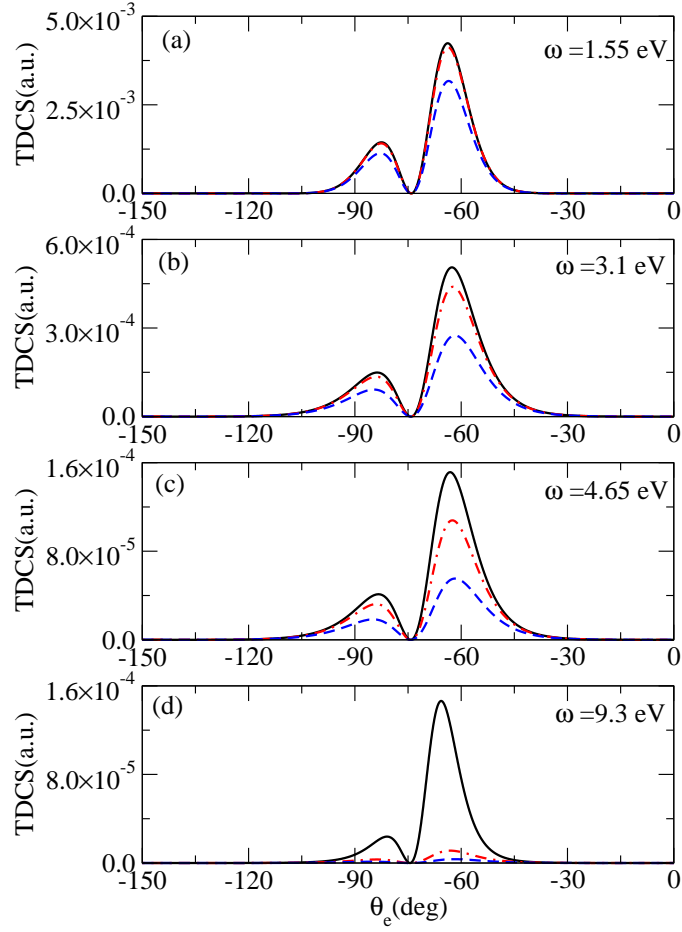}
\caption{(Color online)
Similar result as in Figs. \ref{fig2}(a)-\ref{fig2}(d), but for a scattering angle  
$\theta_f = 15^{\circ}$.}
\label{fig3}
\end{figure}

\noindent
In Fig. \ref{fig2} we present the TDCSs as a function of the angle of the ejected 
electron, $\theta_e$,  with exchange of one photon, $N = 1$, at high kinetic energies of 
the projectile electron $E_{i}=2$ keV and ejected electron $E_e=200$  eV, and a small 
scattering angle, $\theta_f = 5^{\circ}$.
The laser intensity is  $I=1$ TW/cm$^2$, while the photon energies  we consider are: 
$1.55 $ eV in  Fig. \ref{fig2}(a), $ 3.1$ eV in Fig. \ref{fig2}(b), $4.65 $ eV in   Fig. 
\ref{fig2}(c), and $9.3$ eV in Fig. \ref{fig2}(d).
Figure \ref{fig3} show similar results to Fig. \ref{fig2}, but for a larger scattering 
angle $\theta_f = 15^{\circ}$.
In all figures the solid lines correspond to the laser-assisted TDCSs calculated from 
Eq. (\ref{tdcs}), which include the laser dressing effects of the projectile and of the 
hydrogen atom, the dot-dashed lines correspond to the TDCSs in which the atomic dressing 
is considered in the LFA for the direct and as well as exchange scattering, while the 
dashed lines correspond to the results  in which the atomic dressing is neglected.
\noindent
As resulted from our theoretical calculations, the TDCS is quite important at scattering 
and ejected angles where the recoil momentum $q$ is small.
Thus, at the scattering angle $\theta_f = 5^{\circ}$ the angular distribution of the 
electrons is observed with a highest probability at the maximum values of TDCSs, which 
occur  at the following detection angles $\theta_e \simeq -61^\circ $ in Fig. 
\ref{fig2}(a), $\theta_e \simeq -40^\circ $ in Fig. \ref{fig2}(b), $\theta_e \simeq 
-38^\circ $ in Fig. \ref{fig2}(c), and $\theta_e \simeq -39^\circ $ in Fig. \ref{fig2}(d).
Similar to  the free-free transitions or other laser-assisted processes 
\cite{ehl1998,acgabi99,joa2012},  the net effect of the laser field is to decrease 
the peak values of the angular distributions of TDCSs, while the atomic dressing 
contribution is increasing with photon energy.
The dressing effect of the laser is included in the argument of the Bessel 
function through the quiver motion amplitude, $\alpha_0 $, in the electronic transition 
amplitudes, Eqs. (\ref{tne}) and  (\ref{tne_ex}), as well through ${\cal R}_q$ and the 
factor $\alpha_0 \, \omega$ in the atomic transition amplitudes, Eqs. (\ref{tn1}) and 
(\ref{tn1_ex}).
Thus, as the photon energy increases the atomic dressing effects (included in the full 
lines) are more important than the electronic dressing effects (included in the dashed 
lines), and the TDCS decreases as suggested by Eqs. (\ref{tne_CA}), (\ref{tn1lim_CA}),  
and (\ref{ratio_tne_CA}) derived in the low-photon energy limit.
At low photon energies which are far from any atomic resonance, of $1.55 $ 
eV or even $3.1$ eV at $\theta_f = 5^{\circ}$, the laser-assisted $(e,2e)$ process 
depicted in Figs. \ref{fig2}(a)-\ref{fig2}(b) and \ref{fig3}(a) is well described by the 
LFA (dot-dashed lines), as long as the photon energy is much smaller than $|E_1|$.
A clear signature of the nonperturbative effect of the laser is the oscillatory 
character of the angular distribution of TDCS, as shown in Fig. \ref{fig2}(a) compared to 
 Figs. \ref{fig2}(b)-\ref{fig2}(d).
The nonperturbative behavior, due to a larger quiver amplitude, is seen  at the small 
photon energy of $1.55 $ eV  ($\alpha_{0} \simeq 1.64$ a.u. and $U_p \simeq 0.03$ 
eV), and resides in the occurrence of an increasing number of zeros in the Bessel 
functions of the first kind and, therefore, in the TDCS \cite{cionga93}.
Thus, kinematic minima of TDCSs appear at ${\cal R}_q = 0$ if the scalar product 
$\bm{\varepsilon}
\cdot \mathbf{q} =0$, condition that is fulfilled at  ejected electron angles
given by the relation $\cos \theta_e  = {(k_i -k_f\cos \theta_f)}/{k_e}$.
The first two kinematical minima which are located on the left- and right-side of the 
main maximum in Fig. \ref{fig2}(a) at the photon energy  of $1.55 $ eV, occur at the 
angles $\theta_e \simeq  - 79^{\circ}$ and $ 80^{\circ}$, while the next minima of TDCS  
are due to the zeros of the Bessel function $J_1({\cal R}_q)$ with ${\cal R}_q \neq 0$.
At larger photon energies  in Figs. \ref{fig2}(b)-\ref{fig2}(d) the 
first two kinematical minima of the TDCSs occur at the  angles $\theta_e \simeq - 
79^{\circ}$ in Figs. \ref{fig2}(b)-\ref{fig2}(d), and at $ \theta_e \simeq\ 80^{\circ}$ 
in Fig. \ref{fig2}(b), $ \theta_e \simeq\ 81^{\circ}$ in Fig. \ref{fig2}(c), and $ 
\theta_e \simeq\ 83^{\circ}$ in Fig. \ref{fig2}(d).

\begin{figure}
\centering
\includegraphics[width=3.5in,angle=0]{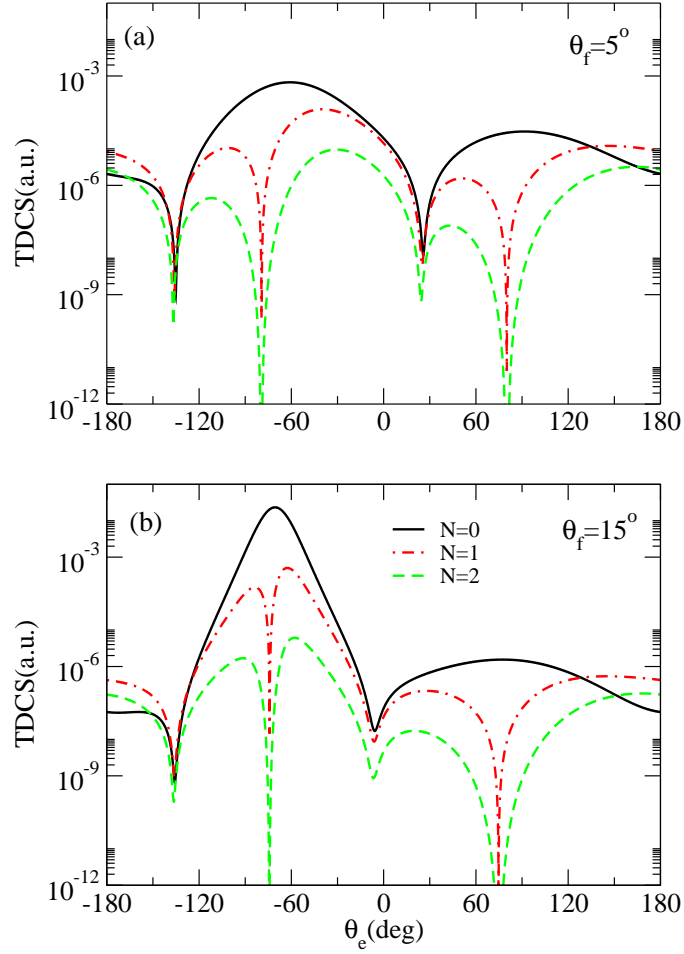}
\caption{(Color online)
The TDCSs for the ionization of hydrogen by electron impact in the presence of a LP laser 
field as a function of the ejected electron angle, $\theta_e$, at the photon energy 
$\omega=3.1 $ eV for no photon exchange (full line), one- and two-photon absorption 
(dot-dashed and dashed lines).
The scattering angle is $\theta_f = 5^{\circ}$ in Fig. \ref{fig4}(a) and $\theta_f = 
15^{\circ}$ in  Fig. \ref{fig4}(b).
The other parameters concerning the scattering geometry, incident and ejected electron 
energies, and laser intensity are the same as in  Fig. \ref{fig2}.}
\label{fig4}
\end{figure}

In Fig. \ref{fig4} we show the numerical results for TDCSs plotted in a logarithmic scale 
for $N=0, 1$ and $2$, at $\omega =3.1$ eV and the scattering angles $\theta_f = 
5^{\circ}$ in Fig. \ref{fig4}(a) and  $\theta_f = 15^{\circ}$ in Fig. \ref{fig4}(b), with 
the same parameters as in  Figs. \ref{fig2}(b) and \ref{fig3}(b).
The angular distributions of TDCS at different $N$ present similar features with 
different magnitudes, and show that the net effect of the laser field  is to decrease the 
values of TDCSs and to split the peaks which occur at $N=0$ (full lines) at $\theta_e 
\simeq  - 62^{\circ}$ in  Fig. \ref{fig4}(a) and $\theta_e \simeq  - 71^{\circ}$ in  Fig. 
\ref{fig4}(b).
The splitting of the peaks by the kinematical minima, which is a well known signature of 
the laser field on the TDCSs, appears due to cancellation of the scalar product 
$\bm{\varepsilon}\cdot \mathbf{q}$ and is located almost symmetrically with respect to 
the direction of the incident electron  at the ejected angles $\theta_e \simeq  - 
79^{\circ}$ and $80^{\circ}$ in Fig. \ref{fig4}(a) and $\theta_e \simeq  - 74^{\circ}$ 
and $75^{\circ}$ in Fig. \ref{fig4}(b). 

\begin{figure}
\centering
\includegraphics[width=3.5in,angle=0]{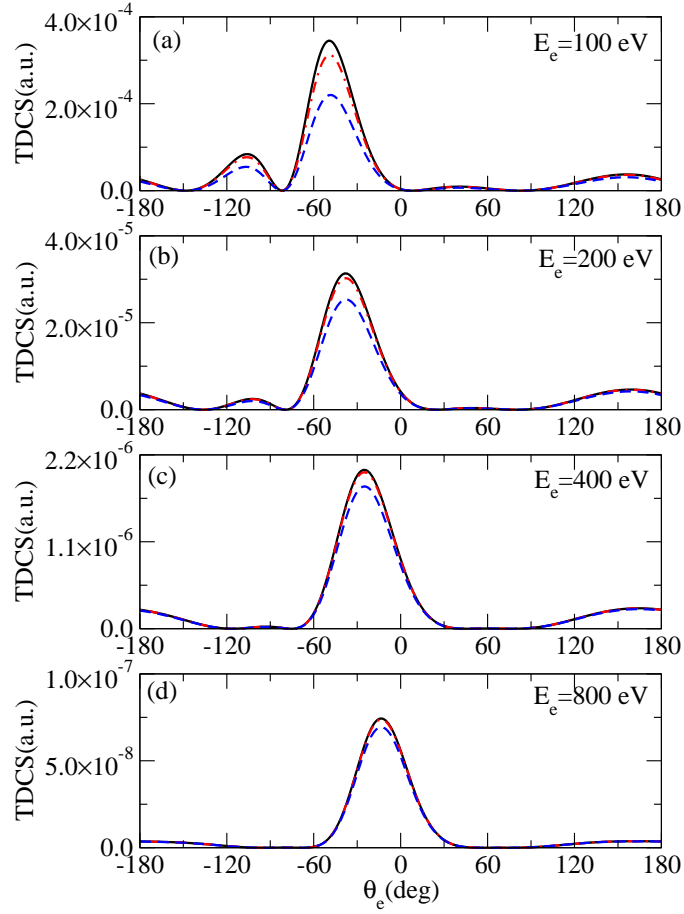}
\caption{(Color online)
The TDCSs for the ionization of hydrogen by electron impact in the presence of a LP laser 
field as a function of the ejected electron angle $\theta_e$, for one-photon absorption.
The photon energy is $\omega=4.65 $ eV, while the  kinetic energy of the ejected electron 
is $E_e=100$ eV in  Fig. \ref{fig5}(a), $ 200$ eV in Fig. \ref{fig5}(b), $400 $ eV in 
Fig. \ref{fig5}(c), and $800$ eV in Fig. \ref{fig5}(d).
The other parameters concerning the scattering geometry, kinetic energy of the incident 
projectile, angle of the scattered electron, and laser intensity are the same as in  Fig. 
\ref{fig2}.
}
\label{fig5}
\end{figure}

\begin{figure}
\centering
\includegraphics[width=3.5in,angle=0]{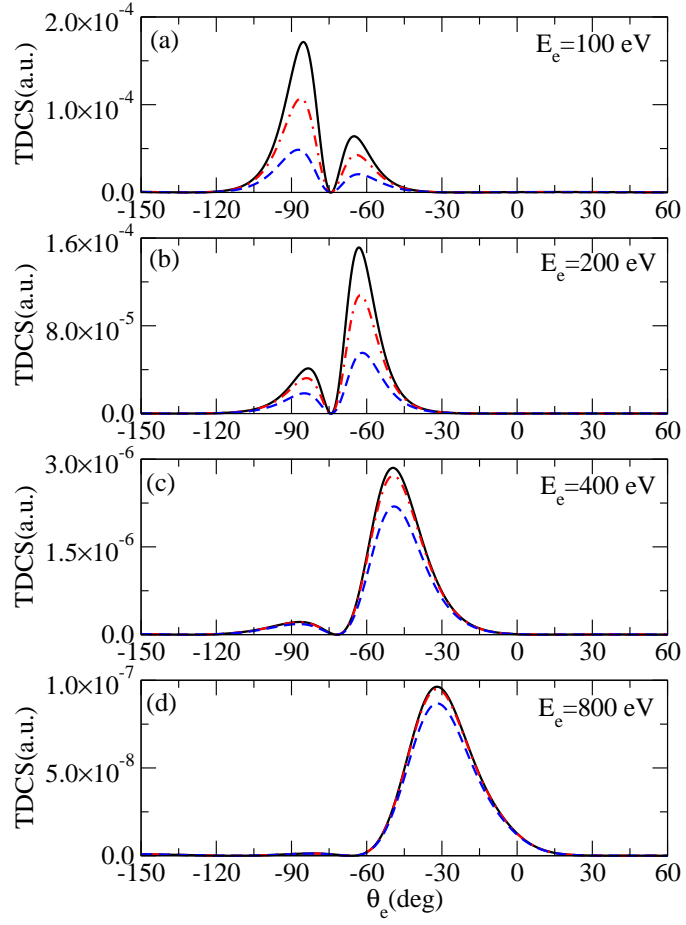}
\caption{(Color online)
Similar result as in Figs. \ref{fig5}(a)-\ref{fig5}(d), but for a scattering angle  
$\theta_f = 15^{\circ}$.}
\label{fig6}
\end{figure}

It is well known the projectile electron plays a major role in the scattering process 
since it interacts with the atomic target electron (a repulsive interaction), its nucleus 
(an attractive interaction), as well as with the laser field.
In Fig. \ref{fig5} we present the TDCSs for the ionization of hydrogen by electron impact 
in the presence of a LP laser field, for absorption of one photon $N=1$, at a photon 
energy of $4.65$ eV as a function of the ejected electron angle. 
The kinetic energy of the ejected electron is $E_e=100$ eV in  Fig. \ref{fig5}(a), $200$ 
eV in Fig. \ref{fig5}(b), $400$ eV in Fig. \ref{fig5}(c), and $800$ eV in Fig. 
\ref{fig5}(d).
The other parameters concerning the scattering geometry, incident projectile energy, 
angle of the scattered electron, and laser field intensity are the same as in Fig. 
\ref{fig2}.
Figure \ref{fig6} shows similar results as in Fig. \ref{fig5}, but for a larger 
scattering angle $\theta_f = 15^{\circ}$. Clearly, the electronic contribution (dashed 
lines) underestimates the angular distribution of TDCS.
 For  kinetic energies of the ejected electron $E_e \le 100$ eV at $\theta_f = 
5^{\circ}$ and $E_e \le 200$ eV at $\theta_f =15^{\circ}$, the atomic dressing effects 
are quite important, and the TDCS calculated in the LFA (dot-dashed 
lines) fails to describe accurately the laser-assisted $(e,2e)$ process.
As the kinetic energy of the ejected electron increases to $800$ eV, at small scattering 
angles the atomic dressing effects are less important than the electronic 
dressing effects, as it is shown in  Figs. \ref{fig5}(d) and  \ref{fig6}(d).
As we approach the symmetric coplanar case of scattered and ejected electrons of 
equal energies, $E_f \simeq E_e  \simeq (E_i+E_1+\omega)/2$, the minimum of the recoil
momentum amplitude $q$ occurs now at larger angles close to  $\theta_f 
\simeq -\theta_e \simeq 45^{\circ}$, Eq. (\ref{q}).

\begin{figure}
\centering
\includegraphics[width=3.5in,angle=0]{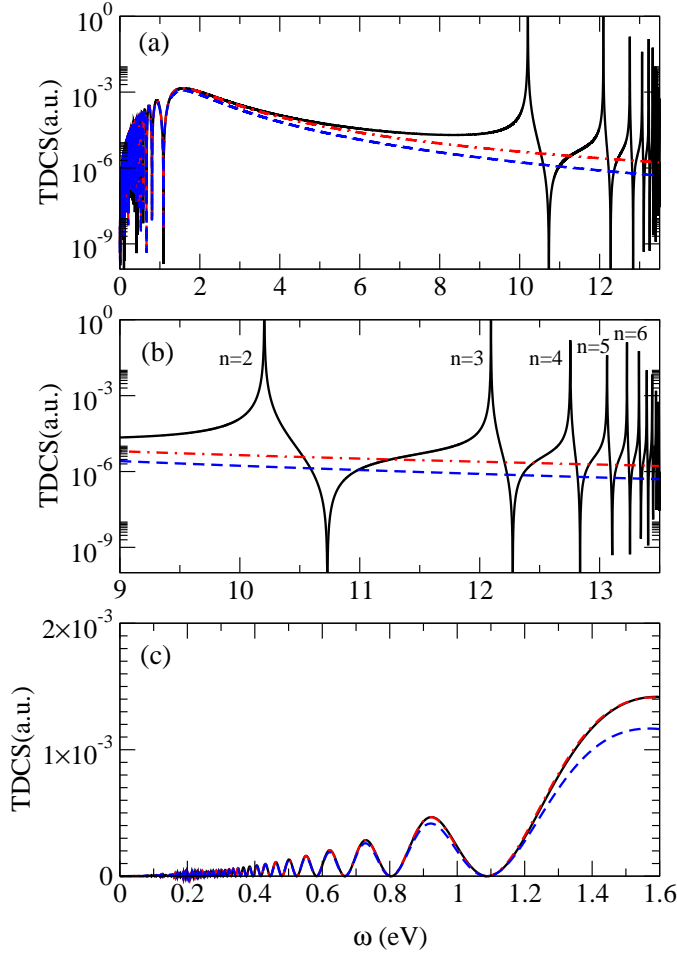}
\caption{(Color online)
The TDCSs for the ionization of hydrogen by electron impact in the presence of a LP laser 
field as a function of the photon energy $\omega$, for one-photon absorption.
The solid lines correspond to the laser-assisted TDCSs calculated from Eq. (\ref{tdcs}), 
the dashed lines correspond to TDCSs in which the atomic dressing is neglected, and 
dot-dashed lines correspond to TDCSs in which the atomic dressing is considered in the 
LFA.
The angles of the scattered and ejected electrons are $\theta_f = 15^\circ$ and $\theta_e 
= -55^\circ$, while the other parameters concerning the scattering geometry, incident and 
ejected electron energies, and laser intensity are the same as in  Fig. \ref{fig2}.
Figure \ref{fig7}(b) shows the detailed structure of the resonance peaks which occur 
at photon energies $\omega= |E_1|(1-1/n^2)$, with $n\ge 2$.
Figure \ref{fig7}(c) shows, in a linear scale, the detailed structure of the oscillatory 
behavior of TDCS at small photon energies, $\omega \le 1.6 $ eV.
}
\label{fig7}
\end{figure}

In order clarify the importance of the atomic dressing term we illustrate in Fig. 
\ref{fig7}(a) the TDCS, in a logarithmic scale, with respect to the photon energy for 
one-photon absorption. 
The kinetic energies of the projectile and the ejected electrons are $E_i = 2$ keV and 
$E_e = 200$ eV, while the angles of the scattered and ejected electrons are chosen 
$\theta_f = 15^\circ$ and $\theta_e = -55^\circ$.
The polarization vector of the electric field is parallel to the momentum of the 
incident electron, and we consider a moderate laser intensity, $I=1 $ TW/cm$^2$, for which 
the non-perturbative dressing effects of the projectile and 
ejected electrons can be visualized at small photon energies with $\alpha_0 > 1$. 
The solid line corresponds to the laser-assisted TDCS calculated from Eq. (\ref{tdcs}), 
which includes the dressing effects of the projectile and of the atomic target, the 
dashed line corresponds to TDCS  in which the atomic dressing terms are 
neglected, while the dot-dashed line corresponds to the result in which the atomic 
dressing terms are considered in the LFA,  Eq. (\ref{tn1lim-LP0}).
 The TDCS  shows a strongly dependence on the atomic structure of the target and exhibits 
a series of resonance peaks which are associated with  one-photon absorption from the 
initial ground state of the hydrogen atom, at photon energies that match atomic 
resonances $\omega= E_n-E_1$, and correspond to poles that occur in the atomic radial 
integral at $\tau =n$, with $n \ge 2$, as detailed in Fig. \ref{fig7}(b).
The noteworthy feature of the atomic radial integral $J_{101}$, Eq. (\ref{J101}), is 
that it presents poles with respect to $\tau$ which arise due to the cancellation of the 
$2- \tau$ factor in the denominator, as well as from the poles of the Appell's 
hypergeometric functions $F_1$ at $\tau=n'$, where $n' \ge 3$ is an integer. The origin of 
these poles resides in the poles of the Coulomb Green's functions used for the calculation 
of the linear-response vector ${\bf w}_{100}$, \cite{vf1}.
Clearly, the LFA, which does not take into account the atomic structure, fails to describe 
the laser-assisted $(e,2e)$ process at large photon energies (typically in the UV range).
Figure \ref{fig7}(c) shows the energy spectra in the nonperturbative regime at low photon 
energies, $\omega \le 1.6$ eV in Fig. \ref{fig7}(a), domain where at the laser 
intensity $I=1$ TW/cm$^2$ the quiver amplitude $\alpha_{0}$ is larger than $1$ a.u. and 
increases up to $395$ a.u. for $\omega=0.1$ eV. 
The TDCS presents oscillations due to the Bessel function $J_1({\cal R}_q)$, and the LFA 
(dot-dashed line) gives a good description of the atomic dressing effect.
It should be kept in mind that both the laser intensity and photon energy  play an 
important role, and obviously, the nonperturbative effects are seen to be important as 
laser intensity increases and photon energy decreases due to increasing quiver motion 
of the free and bound electrons, and contribute to the oscillatory behavior  of 
the laser-assisted TDCSs, which resides in the occurrence of increasing oscillations of 
the Bessel function.

\section{Summary and conclusions}
\label{IV}

We study the electron-impact ionization of hydrogen at large projectile and ejected 
electron kinetic energies in the presence of a linearly polarized laser field, and
 investigate the laser-assisted $(e, 2e)$ reaction at moderate laser field intensities.
We focus our numerical results on the case of the asymmetric coplanar scattering 
geometry, where 
we discuss the importance of the dressing effects and  analyze the influence of the 
laser field on the TDCS in several numerical examples.
The laser-assisted $(e, 2e)$ reaction has a nonlinear character which consists in 
multiphoton absorption (emission) of photons from (to) laser radiation by projectile and 
ejected electrons and  atomic target.
We present a new method to calculate the atomic radial amplitude in a closed form, 
which represents the main difficulty in the evaluation of TDCS.
Thus, a semi-perturbative approach is used, in which for the interaction of the fast 
incident and outgoing electrons with the laser field we employ non-perturbative 
Gordon-Volkov wave functions, while the interaction of the hydrogen atom with the laser 
field is  considered in first-order TDPT, and the interaction of the fast incident 
electron with the hydrogen atom is treated  in the first-order Born approximation.
The exchange between the outgoing electrons can not be ignored when ejected electrons 
with large kinetic energy are detected, and is included in the calculation.
Our theoretical formulas and numerical results clearly demonstrate the strong influence of 
the photon energy and laser intensity on the dynamics of laser-assisted $(e, 2e)$ 
process.
\noindent
It was found that the atomic dressing contribution calculated in first-order TDPT in the 
laser field substantially modifies the laser-assisted TDCSs at small momenta $q$, $ 
\Delta$, and $ \Delta_e$, and for photon energies close to resonances.
The introduction of the laser field in the $(e,2e)$ reaction changes the profile of TDCS 
as it is seen in Fig. \ref{fig4}, where the  peaks of TDCSs  are reduced in magnitude and 
splitted by the presence of the laser, due to appearance of the kinematical minima.
We show that the atomic dressing effects strongly depend on the structure of the atomic 
target as is seen in Fig. \ref{fig7}, and cannot be correctly described by the LFA at 
large photon energies.
At low photon energies we confirm the validity of LFA by comparing the numerical 
results obtained for the atomic matrix elements within LFA with the results obtained by 
first-order TDPT.
\noindent
Thus, the theoretical studies remain very useful for understanding essential 
details of the scattering signal due to the fact that the derived analytical formulas have 
the advantage of giving more physical insight into the laser-assisted $(e, 2e)$ process 
and valuable information in future theoretical and experimental investigations.

\section*{ACKNOWLEDGMENTS}
The work by G. B. was supported by the research program PN 19 15 01 02 through
Contract No. 4N/2019 (Laplas VI) from the UEFISCDI and the Ministry of Research, 
Innovation, and Digitization of Romania.

\clearpage
\newpage

\end{document}